%% file: main.tex
\documentclass[letterpaper,twocolumn,10pt]{article}
\usepackage{usenix,epsfig,endnotes}
\usepackage{listings}
\usepackage{amsfonts}
\usepackage{float}
\usepackage{amssymb}
\usepackage{amscd}
\usepackage{amsmath}
\usepackage{microtype}
\usepackage{array}
\usepackage{graphics}
\usepackage{multirow}
\usepackage{graphicx}
\usepackage[font=small,labelfont=bf]{caption}
\usepackage{pifont}
\usepackage{listings}
\usepackage{subfigure}
\usepackage{booktabs}
\usepackage[title]{appendix}
\usepackage{enumitem,amssymb}
\newlist{checkbox}{itemize}{1}
\setlist[checkbox]{label=$\square$,topsep=2pt,itemsep=1pt,parsep=1pt}
\usepackage{color}
\definecolor{gray}{rgb}{0.4,0.4,0.4}
\definecolor{darkblue}{rgb}{0.0,0.0,0.6}
\definecolor{cyan}{rgb}{0.0,0.6,0.6}

\usepackage[obeyspaces]{url}

\usepackage[
  breaklinks = true,
  unicode = true,
  urlcolor = blue,
  colorlinks = true,
  citecolor = blue,
  linkcolor = red]{hyperref}

\usepackage{authblk}

\newcommand{\ignore}[1]{}
\newcommand{\revised}[1]{}
\newcommand\comment[1]{}

\newcommand\todo[1]{\{\textbf{Todo:} {\em#1}\}}

\newcolumntype{L}[1]{>{\raggedright\let\newline\\\arraybackslash\hspace{0pt}}m{#1}}
\newcolumntype{C}[1]{>{\centering\let\newline\\\arraybackslash\hspace{0pt}}m{#1}}
\newcolumntype{R}[1]{>{\raggedleft\let\newline\\\arraybackslash\hspace{0pt}}m{#1}}

\begin{document}

%don't want date printed
\date{}

%make title bold and 14 pt font (Latex default is non-bold, 16 pt)
%\title{\Large \bf Dangerous Skills: Security Risks of Voice-Controlled Third-Party Functionalities on Virtual Personal Assistant Systems\thanks{All the squatting and impersonation vulnerabilities we discovered are reported to Amazon and Google and received their acknowledgement~\cite{demo}.}}

\title{\Large \bf Understanding and Mitigating the Security Risks of Voice-Controlled Third-Party Skills on Amazon Alexa and Google Home\thanks{All the squatting and impersonation vulnerabilities we discovered are reported to Amazon and Google and received their acknowledgement~\cite{demo}.}}

%\title{\Large \bf Hearing is Deceiving: Security Analysis of Voice-Controlled Virtual Personal Assistant Systems\thanks{All the squatting and impersonation vulnerabilities we discovered are reported to Amazon and Google and received their acknowledgement~\cite{demo}.}}

\author[1]{Nan Zhang}
\author[1]{Xianghang Mi}
\author[1,2]{Xuan Feng}
\author[1]{XiaoFeng Wang}
\author[3]{Yuan Tian}
\author[1]{Feng Qian}
\affil[1]{\small Indiana University, Bloomington \authorcr Email: {\tt \{nz3, xmi, xw7, fengqian\}@indiana.edu}}
\affil[2]{\small Beijing Key Laboratory of IoT Information Security Technology, Institute of Information Engineering, CAS, China \authorcr Email: {\tt fengxuan@iie.ac.cn}}
\affil[3]{\small University of Virginia \authorcr Email: {\tt yuant@virginia.edu}}

\maketitle

% Use the following at camera-ready time to suppress page numbers.
% Comment it out when you first submit the paper for review.
\thispagestyle{empty}

\subsection*{Abstract}

Virtual personal assistants (VPA) (e.g., Amazon Alexa and Google Assistant) today mostly rely on the voice channel to communicate with their users, which however is known to be vulnerable, lacking proper authentication. \ignore{Prior research shows that obfuscated voice commands and inaudible ultrasound can be used to attack these systems. }The rapid growth of VPA skill markets opens a new attack avenue, potentially allowing a \textit{remote} adversary to publish attack skills to attack a large number of VPA users through popular IoT devices such as Amazon Echo and Google Home. In this paper, we report a study that concludes such remote, large-scale attacks are indeed realistic. More specifically, we implemented two new attacks: \textit{voice squatting} in which the adversary exploits the way a skill is invoked (e.g., ``open capital one''), using a malicious skill with similarly pronounced name (e.g., ``capital won'') or paraphrased name (e.g., ``capital one please'') to hijack the voice command meant for a different skill, and \textit{voice masquerading} in which a malicious skill impersonates the VPA service or a legitimate skill to steal the user's data or eavesdrop on her conversations.  These attacks aim at the way VPAs work or the user's misconceptions about their functionalities, and are found to pose a realistic threat by our experiments (including user studies and real-world deployments) on Amazon Echo and Google Home. The significance of our findings have already been acknowledged by Amazon and Google, and further evidenced by the risky skills discovered on Alexa and Google markets by the new detection systems we built. We further developed techniques for automatic detection of these attacks, which already capture real-world skills likely to pose such threats.

%With virtual personal assistant (VPA) systems (Amazon Alexa, Google assistant) and their IoT devices they power (e.g., Amazon Echo, Google Home) gaining popularity,

\input{1_introduction}

\input{2_background}

\input{3_attack}

\input{4_measurement}

\input{5_defense}

\input{7_relatedwork}

\input{8_conclusion}

{\footnotesize \bibliographystyle{acm}
\bibliography{ref}}

\input{9_appendix}

%\theendnotes

\end{document}

%% file: 1_introduction.tex
\section{Introduction}
\label{sec:introduction}

The wave of Internet of Things (IoT) has brought in a new type of \textit{virtual personal assistant} (VPA) services.  Such a service is typically delivered through a smart speaker that interacts with the user using a voice user interface (VUI), allowing the user to command the system with voice only: for example, one can say ``what will the weather be like tomorrow?'' ``set an alarm for 7 am tomorrow'', etc., to get the answer or execute corresponding tasks on the system. In addition to their built-in functionalities, VPA services are enhanced by \textit{ecosystems} fostered by their providers, such as Amazon and Google, under which third-party developers can build new applications (called \textit{skills} by Amazon and \textit{actions} by Google\footnote{Throughout the paper, we use the Amazon term \textit{skill} to describe third-party applications, including Google's actions.}) to offer further helps to the end users, for example, order food, manage bank accounts and text friends. In the past year, these ecosystems are expanding at a breathtaking pace: Amazon claims that already 25,000 skills have been uploaded to its skill market to support its VPA system (including the \textit{Alexa} VPA service running through Amazon Echo)~\cite{amazon_25000} and Google also has more than one thousand actions available on its market for its Google Home system (powered by \textit{Google Assistant}). These systems have already been deployed to the households around the world, and utilized by tens of millions of users. This quickly-gained popularity, however, could bring in new security and privacy risks, whose implications have not been adequately investigated so far.

\vspace{3pt}\noindent\textbf{Security risks in VPA voice control}. As mentioned earlier, today's VPA systems are designed to be primarily commanded by voice. \textit{Protecting such VUIs is fundamentally challenging, due to the lack of effective means to authenticate the parties involved in the open and noisy voice channel}. Already prior research shows that the adversary can generate obfuscated voice commands~\cite{197215} or even completely inaudible ultrasound~\cite{Zhang:2017:DIV:3133956.3134052} to attack speech recognition systems. These attacks exploit unprotected communication to impersonate the user to the voice-controlled system, under the constraint that an attack device is placed close to the target (e.g., in the ultrasound attack, within 1.75 meters).

The emergence of the VPA ecosystem completely changes the game, potentially opening new avenues for \textit{remote attacks}. Through the skill market, an adversary can spread malicious code, which will be silently invoked by voice commands received by a VPA device (e.g., Amazon Echo or Google Home). As a result, the adversary gains (potentially large-scale) access to the VPA devices interacting with victims, allowing him to \textit{impersonate a legitimate application or even the VPA service to them}. Again, the attack is made possible by the absence of effective authentication between the user and the VPA service over the voice channel. Our research shows that such a threat is indeed realistic.

\vspace{3pt}\noindent\textbf{Voice-based remote attacks}. In our research, we analyzed the most popular VPA IoT systems -- Alexa and Google Assistant, focusing on the third-party skills deployed to these devices for interacting with end users over the voice channel. Our study demonstrates that through publishing malicious skills, it is completely feasible for an adversary to remotely attack the users of these popular systems, collecting their private information through their conversations with the systems. More specifically, we identified two threats never known before, called \textit{voice squatting attack} (VSA) and \textit{voice masquerading attack} (VMA). In a VSA, the adversary exploits how a skill is invoked (by a voice command), and the variations in the ways the command is spoken (e.g., phonetic differences caused by accent, courteous expression, etc.) to cause a VPA system to trigger a malicious skill instead of the one the user intends (Section~\ref{subsec:voice_squatting}).  For example, one may say ``Alexa, open Capital One please'', which normally opens the skill \textit{Capital One}, but can trigger a malicious skill \textit{Capital One Please} once it is uploaded to the skill market. A VMA aims at the interactions between the user and the VPA system, which is designed to hand over all voice commands to the currently running skill, including those supposed to be processed by VPA system like stopping the current skill and switching to a new one. In response to the commands, a malicious skill can pretend to yield control to another skill (switch) or the service (stop), yet continue to operate stealthily to impersonate these targets and get sensitive information from the user \ignore{, for example, XXX }(Section~\ref{subsec:voice_masquerading}).

We further investigated the feasibility of these attacks through user studies, system analysis, and real-world exploits. More specifically, we first surveyed 156 Amazon Echo and Google Home users and found that most of them tend to use natural languages with diverse expressions to interact with the devices: e.g., ``play some sleep sounds''. These expressions allow the adversary to mislead the service and launch a wrong skill in response to the user's voice command, such as \textit{some sleep sounds} instead of \textit{sleep sounds}. Our further analysis of both Alexa and Google Assistant demonstrates that indeed these systems identify the skill to invoke by looking for the longest string matched from a voice command (Section~\ref{subsec:voice_squatting}). Also interestingly, our evaluation of both devices reveals that Alexa and Google Assistant cannot accurately recognize some skills' invocation names and the malicious skills carrying similar names (in terms of pronunciation) are capable of hijacking the brands of these vulnerable skills.

Finally, we deployed four skills through the Amazon market to attack a popular Alexa skill ``Sleep and Relaxation Sounds''~\cite{sleep_sounds}. These skills have been invoked by over 2,699 users in a month and collected 21,308 commands. We built the skills in a way to avoid collecting private information of the real-world users. Still, the commands received provide strong evidence that indeed both voice squatting and masquerading can happen in real life: our study shows that the received commands include the ones only eligible for ``Sleep and Relaxation Sounds'', and those for switching to a different skill or stopping the current skill that can be leveraged to impersonate a different skill (Section~\ref{subsec:real_world_attacks}). Our analysis of existing skills susceptible to the threat further indicates the significant consequences of the attacks, including disclosure of one's home address, financial data, etc. The video demos of these attacks are available online~\cite{demo}.

\vspace{3pt}\noindent\textbf{Responsible disclosure}. We have reported our findings to Amazon and Google, both of which acknowledged the importance of the weaknesses we discovered. And we are helping them to understand and mitigate such new security risks.

\vspace{3pt}\noindent\textbf{Mitigation}. In our research, we developed a suite of new techniques to mitigate the realistic threats posed by VSA and VMA. We built a \textit{skill-name scanner} that converts the invocation name string of a skill into a phonetic expression specified by ARPABET~\cite{arpabet}. This expression describes how a name is pronounced, allowing us to measure the phonetic distance between different skill names. Those sounding similar or having a subset relation are automatically detected by the scanner. This technique can be used to vet the skills uploaded to a market. Interestingly, when we ran it against all 19,670 custom skills on the Amazon market, we discovered 4,718 skills with squatting risks. \ignore{Particularly, a shopping skill ``Wali'' apparently is dangerously close to another skill ``Wally'' that controls home automation devices.} These findings indicate that possibly these attacks could already happen in the real world.

To counter the threat of the masquerading attack, we designed and implemented a novel context-sensitive detector to help a VPA service capture the commands for system-level operations (e.g., invoke a skill) and the voice content unrelated to a skill's functionalities, which therefore should not be given to a skill (Section~\ref{sec:defense}). 
Specifically, our detection scheme consists of two components: the \textit{Skill Response Checker (SRC)} and the \textit{User Intention Classifier (UIC)}. SRC captures suspicious skill responses that a malicious skill may craft, such as a fake skill recommendation mimicking that from the VPA system. UIC instead examines the information flow of the opposite direction, i.e., utterances from the user, to accurately identify users' intents of context switches. 
Built upon robust Natural Language Processing (NLP) and machine learning techniques,
SRC and UIC form two lines of defense towards the masquerading attack based on extensive empirical evaluations.

\vspace{3pt}\noindent\textbf{Contributions}. The contributions of the paper are outlined as follows:

%\begin{itemize}[topsep=1pt]
%\itemsep0em
\vspace{2pt}\noindent$\bullet$\textit{ First study on remote VPA attacks}. We report the first security analysis on the VPA ecosystems and related popular IoT devices (Amazon Echo and Google Home), which leads to the discovery of serious security weaknesses in their VUIs and skill vetting. We present two new attacks, voice squatting and voice masquerading. Both are demonstrated to pose realistic threats to a large number of VPA users from remote and both have serious security and privacy implications. Our preliminary analysis of the Amazon skill market further indicates the possibility that similar attacks may already happen in the real world.

\vspace{2pt}\noindent$\bullet$\textit{ New techniques for risk mitigation}.  We made the first step towards protecting VPA users from these voice-based attacks. We show that the new protection works effectively against the threats in realistic environments. The idea behind our techniques, such as context-sensitive command analysis, could inspire further enhancement of the current designs to better protect VPA users.

%\end{itemize}

%\vspace{3pt}\noindent\textbf{Roadmap}.  The rest of the paper is organized as follows: Section~\ref{sec:background} presents the background of the VPA IoT systems and their potential security risks; Section~\ref{sec:attack} elaborates our security analysis of these systems and the two new attacks discovered in our research; Section~\ref{sec:measurement} describes the technique to identify suspicious skill names and our findings from scanning the Amazon market using the technique; Section~\ref{sec:defense} introduces the protection that detects suspicious user-skill conversations; Section~\ref{sec:discussion} discusses the limitations of the study and envisions future research; Section~\ref{sec:relatedwork} reviews related prior work and Section~\ref{sec:conclusion} concludes the paper.

%% file: 2_background.tex
\section{Background}
\label{sec:background}

\subsection{Virtual Personal Assistant Systems}

\ignore{The first modern virtual personal assistant, called \textit{Siri} was introduced to smartphone by Apple in 2011. Later, VPA systems were quickly developed and integrated into many types of platforms, e.g. mobile operating system, IoT devices, instant messaging app.}

\vspace{3pt}\noindent\textbf{VPA on IoT devices}. Amazon and Google are two major players in the market of smart speakers with voice-controlled personal assistant capabilities. Since the debut of the first \textit{Amazon Echo} in 2015, Amazon has now taken 76\% of the U.S. market with an estimate of 15-million devices sold in the U.S. alone in 2017~\cite{amazon_market_share}. In the meantime, Google has made public \textit{Google Home} in 2016, and now grabbed the remaining 24\% market share. \textit{Amazon Echo Dot} and \textit{Google Home Mini} are later released in 2016 and 2017, respectively, as small, affordable alternatives to their more expensive counterparts. Additionally, Amazon has integrated VPA into IoT products from other vendors, e.g. Sonos smart speaker, Ecobee thermostat~\cite{avs}.

A unique property of these four devices is that they all forgo conventional I/O interfaces, such as the touchscreen, and also have fewer buttons (to adjust volume or mute), which serves to offer the user a hands-free experience. In another word, one is supposed to command the device mostly by speaking to it. For this purpose, the device is equipped with a microphone circular array designed for 360-degree audio pickup and other technologies like beam forming\ignore{, noise reduction} that enable far-field voice recognition. Such a design allows the user to talk to the device anywhere inside a room and still get a quick response.

\vspace{3pt}\noindent\textbf{Capabilities}. Behind these smart devices is a virtual personal assistant, called \textit{Alexa} for Amazon and \textit{Google Assistant} for Google, engages users through a two-way conversation. Unlike those serving a smartphone (\textit{Siri}, for example) that can be activated by a button push, the VPAs for these IoT devices are started with a wake-word like ``\textit{Alexa}'' or ``\textit{Hey Google}''. These assistants have a range of capabilities, from weather report, timer setting, to-do list maintenance to voice shopping, hands-free messaging and calling. The user can manage these capabilities through a companion app running on her smartphone. 

\subsection{VPA Skills and Ecosystem}
\label{subsec:vpa_skill}

Both Amazon and Google enrich the VPAs' capabilities by introducing voice assistant \textit{skill} (or \textit{action} on Google). Skills are essentially third-party apps, like those running on smartphones, offering a variety of services the VPA itself does not provide.  Examples include \textit{Amex}, \textit{Hands-Free Calling}, \textit{Nest Thermostat} and \textit{Walmart}. These skills can be conveniently developed with the supports from Amazon and Google, using \textit{Alexa Skills Kit}~\cite{kumar2017just} and \textit{Actions on Google}. Indeed, we found that up to November 2017, Alexa already has 23,758 skills and Google Assistant has 1,001. More importantly, new skills have continuously been added to the market, with their total numbers growing at a rate of 8\% for Alexa and 42\% for Google Assistant, as we measured in a 45-day period.

%\vspace{3pt}\noindent\textbf{Explicit skill invocation}. Skills can be started explicitly when a user requires a skill by its name from a VPA: for example, saying ``Alexa, talk to Amex'' to Alexa triggers the \textit{Amex} skill for making a payment or checking bank account balances. Such a type of skills are also called \textit{custom skills} on Alexa.

\vspace{3pt}\noindent\textbf{Skill invocation}. Skills can be started either explicitly or implicitly. Explicit invocation takes place when a user requires a skill by its name from a VPA: for example, saying ``Alexa, talk to Amex'' to Alexa triggers the \textit{Amex} skill for making a payment or checking bank account balances. Such a type of skills are also called \textit{custom skills} on Alexa. 

Implicit invocation occurs when a user tells the voice assistant to perform some tasks without directly calling to a skill name. For example, ``Hey Google, will it rain tomorrow?'' will invoke the \textit{Weather} skill to respond with a weather forecast. Google Assistant identifies and activates a skill implicitly whenever the conversation with the user is under the context deemed appropriate for the skill. This invocation mode is also supported by the Alexa for specific types of skills\ignore{, for those belonging to \textit{Smart Home Skills}, \textit{Flash Briefing Skills}, \textit{Video Skills}, and \textit{List Skills}}.

\vspace{3pt}\noindent\textbf{Skill interaction model}. The VPA communicates with its users based upon an \textit{interaction model}, which defines a loose protocol for the communication. Using the model, the VPA can interpret each voice request, translating it to the command that can be handled by the VPA or a skill.

Specifically, to invoke a skill explicitly, the user is expected to use a wake-word, a trigger phrase and the skill's invocation name. For example, for the spoken sentence ``Hey Google, talk to personal chef'', ``Hey Google'' is the wake-word, ``talk to'' is the trigger phrase, and ``personal chef'' is the skill invocation name. Here, trigger phrase is given by the VPA system, which often includes the common terms for skill invocation like ``open'', ``ask'', ``tell'', ``start'' etc. Note that skill invocation name could be different from skill name, which is intended to make it simpler and easier for users to pronounce. For example, ``The Dog Feeder'' has invocation name as \textit{the dog}; ``Scryb'' has invocation name as \textit{scribe}.

%Specifically, to invoke a skill explicitly, the user is expected to use a wake-word, a trigger phrase, the skill's invocation name and/or an action phrase. For example, for the spoken sentence ``Hey Google, talk to personal chef to find a cannoli recipe'', ``Hey Google'' is the wake-work, ``talk to'' is the trigger phrase, ``personal chef'' is the skill invocation name, and ``to find a cannoli recipe'' is the action phrase. Here, trigger phrase is given by the VPA system, which often includes the common terms for skill invocation like ``open'', ``ask'', ``tell'', ``start'' etc.; action phrase is optional but when used can be mapped to a specific \textit{intent} (explained below) defined by the skill. Note that skill invocation name could be different from skill name, which is intended to make it simpler and easier for users to pronounce. For example, ``The Dog Feeder'' has invocation name as \textit{the dog}; ``Scryb'' has invocation name as \textit{scribe}.

When developing a skill, one needs to define \textit{intents} and \textit{sample utterances} to map the user's voice inputs to various interfaces of the skill that take the actions the user expects. Such an interface is described by the intent. To link a sentence to an intent, the developer specifies sample utterances, which are essentially a set of sentence templates describing the possible ways the user may talk to the skill. There are also some built-in intents within the model like \texttt{WelcomeIntent}, \texttt{HelpIntent}, \texttt{StopIntent}, etc., which already define many common sample utterances. The developer can add more intent or simply specify one default intent, in which case all user requests will be mapped to this intent.

\begin{figure}
\centering
\includegraphics[width=0.4\textwidth]{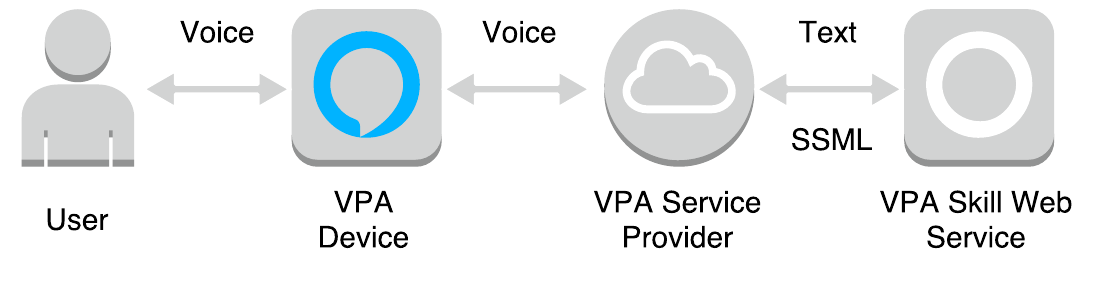}
\caption{Infrastructure of VPA System}
\label{fig:infra}
\vspace{-18pt}
\end{figure}

\vspace{3pt}\noindent\textbf{Skill service and the VPA ecosystem}. A third-party skill is essentially a web service hosted by its developer, with its name registered with the VPA service provider (Amazon or Google), as illustrated in Figure~\ref{fig:infra}. When a user invokes a VPA device with its wake-word, the device captures her voice command and sends it to the VPA service provider's cloud for processing. The cloud performs speech recognition to translate the voice record into text, finds out the skill to be invoked, and then delivers the text, together with the timestamp, device status, and other meta-data, as a request to the skill's web service. In response to the request, the service returns a response whose text content, either in plaintext or in the format of Speech Synthesis Markup Language (SSML)~\cite{ssml}, is converted to speech by the cloud, and played to the user through the device.  SSML also allows the skill to attach audio files (such as MP3) to the response, which is supported by both Amazon and Google.

Both Amazon and Google have skill markets to publish third-party skills. To publish a skill, the developer needs to submit the information about her skill like name, invocation name, description and the endpoint where the skill is hosted for a certification process. This process aims at ensuring that the skill is functional and meets the VPA provider's security requirements and policy guidelines. 

Once a skill is published, users can simply activate it by calling its invocation name. Note that unlike smartphone apps or website plugins that need to be installed by users explicitly, skills can be automatically discovered (according to the user's voice command) and transparently launched directly through IoT devices. 

\subsection{Adversary Model}

We consider the adversary aiming a large-scale remote attack on the VPA users through publishing malicious skills. Such skills can be transparently invoked by the victim through voice commands, without being downloaded and installed on the victim's device. Therefore, they can easily affect a large number of VPA IoT devices. For this purpose, we assume that the adversary has the capability to build the skill and upload it to the market. This can be easily done in practice, as we found in our research. To mitigate this threat, our protection needs to be adopted by the VPA provider, for vetting submitted skills and evaluating the voice commands received. This requires that the VPA service itself is trusted.

%% file: 3_attack.tex
\section{Exploiting VPA Voice Control}
\label{sec:attack}

\subsection{Analysis of VPA Voice Control}
\label{subsec:vpa_analysis}

\noindent\textbf{Security risks of rogue skills}. As mentioned earlier, VPA skills are launched transparently when a user speaks their invocation names (which can be different from their names displayed on the skill market). Surprisingly, we found that for Amazon, such names are not unique skill identifiers: multiple skills with same invocation names are on the Amazon market. Also, skills may have similar or related names. For example, 66 different Alexa skills are called \textit{cat facts}, 5 called \textit{cat fact} and 11 whose invocation names contain the string ``\textit{cat fact}'', e.g. \textit{fun cat facts}, \textit{funny cat facts}. When such a common name is spoken, Alexa chooses one of the skills based on some undisclosed policies (possibly random as observed in our research).  When a different but similar name is called, however, longest string match is used to find the skill. For example,  ``Tell me funny cat facts'' will trigger \textit{funny cat facts} rather than \textit{cat facts}. This problem is less serious for Google, which does not allow duplicated invocation names. However, it also cannot handle similar names. Further discovered in our research is that some invocation names \textit{cannot} be effectively recognized by the speech recognition systems of Amazon and Google. As a result, even a skill with a different name can be mistakenly invoked, when the name is pronounced similarly to that of the intended one.
%This ambiguity in skill identification opens the door for a squatting attack, when the adversary submits a program with an invocation name likely to be confused with that of the victim skill. 
 
Also, we found that the designs of these VPA systems fail to take into full account their users' perceptions about how the systems work. Particularly, both Alexa and Google Assistant run their skills in a simple operation mode in which only one skill executes at a time and it needs to stop before another skill can be launched. However, such a design is not user-friendly and there is no evidence that the user understands that convenient context switch is not supported by these systems. 
%Instead, they are marketed as the systems the user can freely ask any questions. As a result, a malicious skill could introduce a misconception that it can hand over control to a different skill, while in fact impersonating that skill to the user. 

Further, both Alexa and Google Assistant supports volunteer skill termination. For Alexa, the termination command ``Stop'' is delivered to the skill, which is supposed to stop itself accordingly. For Google Assistant, though the user can explicitly terminate a skill by saying ``Stop'', oftentimes the skill is supposed to stop running once its task is accomplished (e.g., reporting the current weather). We found in our research that there is no strong indication whether a skill has indeed quitted. Although Amazon Echo and Google Home have a light indicator, both of which \ignore{are on top of the devices and }will light up when the devices are speaking and listening. However, they could be ignored by the user, particularly when she is not looking at the devices or her sight is blocked when talking.

%Further, both Alexa and Google assistant supports volunteer skill termination. For Alexa, the termination command ``Stop'' is delivered to the skill, which is supposed to stop itself accordingly. For Google assistant, though the user can explicitly terminate a skill by saying ``Stop'', oftentimes the skill is supposed to stop running once its task is accomplished (e.g., reporting the current weather). We found in our research that there is no strong indication whether a skill has indeed quited. Although Amazon Echo is equipped with a ring indicator and Google Home has a light indicator, both of which are on top of the devices and will light up when the devices are speaking and listening. However, they could be ignored by the user, particularly when she is not looking at the devices or her sight is blocked when talking. Indeed, a survey study we conducted shows that \update{69\%} of Alexa and Google Home users do not pay attention to the indicators when talking, and only \update{25\%} of them check whether a conversation ends based on light indicator (see below).

\vspace{3pt}\noindent\textbf{Survey study}. To understand user behaviors and perceptions of voice-controlled VPA systems, which could expose the users to security risks, we surveyed Amazon Echo and Google Home users, focusing on the following questions:

%\begin{itemize}[topsep=1pt]
%\itemsep0em
\vspace{2pt}\noindent$\bullet$ What would you say when invoking a skill?

\vspace{2pt}\noindent$\bullet$ Have you ever invoked a wrong skill?

\vspace{2pt}\noindent$\bullet$ Did you try context switch when talking to a skill?

\vspace{2pt}\noindent$\bullet$ Have you experienced any problem closing a skill?

\vspace{2pt}\noindent$\bullet$ How do you know whether a skill has stopped?
%\end{itemize}

Using Amazon Mechanical Turk, we recruited adult participants who own Amazon Echo or Google Home devices and have used skills before and paid them one dollar for completing the survey\ignore{\footnote{The studies were approved by IRB}}. To ensure that all participants meet the requirements, we asked them to describe several skills and their interactions with the skills \ignore{and the VPA system} and removed those whose answers were deemed irrelevant. In total, we have collected 105 valid responses from Amazon Echo users and 51 valid responses from Google Home users with diverse background (age ranges from 18 to 74 with average age as 37 years; 46\% are female and 54\% are male; education ranges from high school to graduate degree; 21 categories of occupation). On average, each participant reported to have 1.5 devices and used 5.8 skills per week.

%, among which the average age is 37 years; 46\% are female and 54\% are male. On average, each participant reported to have 1.5 devices and used 5.8 skills per week.

In the first part of the survey, we studied how users invoke a skill. For this purpose, we used two popular skills ``Sleep Sounds'', ``Cat Facts'' (``Facts about Sloths'' on Google Home), and let the participants choose the invocation utterances they tend to use for launching these skills (e.g., ``open \textit{Sleep Sounds} please'') and required them to provide additional examples. We then asked the participants whether they ever triggered a wrong skill\ignore{ and to share the experience if they did}. In the following part of the survey, we tried to find out whether the participants attempted to switch context when interacting with a skill, that is, invoking a different skill or directly talking to the VPA service (e.g., adjusting volume). The last part of the survey was designed to study the user experience in stopping the current skill, including the termination utterances they tend to use, troubles they encountered during the termination process and importantly, the indicator they used to determine whether the skill has stopped. Sample survey questions are listed in Appendix ~\ref{appendix:survey_question}.

\begin{table}
\scriptsize
\centering
\caption{Survey responses of Amazon Echo and Google Home users}
\vspace{-5pt}
\label{table:survey}
\begin{tabular}{R{4.8cm} c c}
\toprule
  &  \textbf{Amazon} & \textbf{Google} \\ \toprule
  \multicolumn{1}{L{5cm}}{\textbf{Invoke a skill with natural sentences:}} & &  \\
  Yes, ``open \textit{Sleep Sounds} please'' & 64\% & 55\% \\
  Yes, ``open \textit{Sleep Sounds} for me'' & 30\% & 25\% \\
  Yes, ``open \textit{Sleep Sounds} app'' & 26\% & 20\% \\
  Yes, ``open my \textit{Sleep Sounds}'' & 29\% & 20\% \\
  Yes, ``open the \textit{Sleep Sounds}'' & 20\% & 14\% \\
  Yes, ``play some \textit{Sleep Sounds}'' & 42\% & 35\% \\
  Yes, ``tell me a \textit{Cat Facts}'' & 36\% & 24\% \\
  No, ``open \textit{Sleep Sounds}'' & 13\% & 14\% \\ \midrule
  
  \multicolumn{1}{L{5cm}}{\textbf{Invoke a skill that did not intend to:}} & &  \\
  Yes & 29\% & 27\% \\
  No & 71\% & 73\% \\ \midrule
  
  \multicolumn{1}{L{5cm}}{\textbf{Tried to invoke a skill while interacting with another skill:}} & &  \\
  Yes & 26\% & 24\% \\
  No & 74\% & 76\% \\ \midrule
  
  \multicolumn{1}{L{5cm}}{\textbf{Tried to adjust volume by voice while interacting with another skill:}} & &  \\
  Yes & 48\% & 51\% \\
  No & 52\% & 49\% \\ \midrule
  
  \multicolumn{1}{L{5cm}}{\textbf{Unsuccessful quitting a skill:}} & &  \\
  Yes & 30\% & 29\% \\
  No & 70\% & 71\% \\ \midrule
  
  \multicolumn{1}{L{5cm}}{\textbf{Indicator of the end of a conversation:}} & &  \\
  VPA says ``Goodbye'' or something similar & 23\% & 37\% \\
  VPA does not talk anymore & 52\% & 45\% \\ 
  The light on VPA device is off & 25\% & 18\% \\ \toprule
\end{tabular}
\vspace{-18pt}
\end{table}

Table~\ref{table:survey} summarizes the responses from both Amazon Echo and Google Home users. The results show that more than 85\% of them tend to use natural utterances to open a skill (e.g., ``open \textit{Sleep Sounds} please''), instead of the standard one (like ``open \textit{Sleep Sounds}''). This indicates that it is completely realistic for the user to launch a wrong skill whose name is better matched to the utterances than that of the intended skill (e.g., \textit{Sleep Sounds}). Indeed, 28\% users reported that they did open unintended skills when talking to their devices.

Also interestingly, our survey shows that nearly half of the participants tried to switch to another skill or to the VPA service (e.g. adjusting volume) when interacting with a skill. Such an attempt failed since such context switch is neither supported by Alexa nor Google Assistant. However, it is imaginable that a malicious skill receiving such voice commands could take advantage of this opportunity to impersonate the skill the user wants to run, or even the VPA service (e.g., cheating the user into disclosing personal information for executing commands). Finally, 30\% of the participants were found to experience troubles in skill termination and 78\% did not use the light indicators on the devices as the primary indicator of skill termination. Again, the study demonstrates the feasibility of a malicious skill to fake its termination and stealthily collect the user's information.   

%users rely on other indicators to tell when a skill stops rather than the light indicator, which suggesting that users are not clear about the termination of a skill.

Following we show how the adversary can exploit the gap between the user perception and the real operations of the system to launch voice squatting and masquerading attacks. 

%\vspace {5pt}\noindent\textbf{Attack overview}. We demonstrate, through a user study, that user had problems invoking a skill and prefer to use more natural utterances to invoke a skill. A malicious skill developer could trick users into invoking their skills by squatting an invocation name similar to what users said, which will be demonstrated in section~\ref{}. On the other hand, some users are unaware of incapability of context switching and termination of a skill, which could be deceived by a malicious skill who pretends to satisfy such user requests and impersonate the VPA system or other skills. We will demonstrate such attacks in section~\ref{}.

\subsection{Voice Squatting}
\label{subsec:voice_squatting}

\noindent\textbf{Invocation confusion}. As mentioned earlier, a skill is triggered by its invocation name, which is supposed to be unambiguous and easy to recognize by the devices. Both Amazon and Google suggests that skill developers test invocation names and ensure that their skills can be launched with a high success rate. However, we found that an adversary can intentionally induce confusion by using the name or similar one of a target skill, to trick the user into invoking an attack skill when trying to open the target.
\ignore{Amazon allows two different skills to have the same invocation name, and when the name is spoken, one of them is triggered depending on undisclosed policies. Google disallows duplicated names but cannot identifies those that look different and sound similar.}
For example, the adversary who aims at \textit{Capital One} could register a skill \textit{Capital Won}, \textit{Capitol One}, or \textit{Captain One}. All such names when spoken by the user could become less distinguishable, particularly in the presence of noise, due to the limitations of today's speech recognition techniques. 

Also, this voice squatting attack can easily exploit the longest string match strategy of today's VPAs, as mentioned earlier. Based on our user survey study, around 60\% of Alexa and Google Home users have used the word ``please'' when launching a skill, and 26\% of them attach ``my'' before the skill's invocation name. So, the adversary can register the skills like \textit{Capital One Please} to hijack the invocation command meant for \textit{Capital One}. 
%once an attack has registered a skill with the victim invocation name and the extra words, users who invoked the skill using that extra words will invoke the attacker's skill instead. This is due to the fact that VPA systems of Alexa and Google assistant uses longest matching string to determine which skill to be invoked. For example, ``open \textit{Capital One} please'' would unanimously invoke the \textit{Capital One} skill, until an attacker registered \textit{Capital One Please} skill and in which case, \textit{Capital One Please} will be invoked instead. 

Note that to make it less suspicious, homophones or words pronounced similarly can be used here, e.g. \textit{Capital One Police}\ignore{, ``\textit{scryb}'' for ``\textit{scribe}''}. Again, this approach defeats Google's skill vetting, allowing the adversary to publish the skill with an invocation name unique in spelling but still confusing (with a different skill) in pronunciation. 

To find out whether such squatting attacks can evade skill vetting, we registered 5 skills with Amazon and 1 with Google. These skills' invocation names and the target's name are shown in Table~\ref{table:skills}. All these skills passed the Amazon and Google's vetting process, which suggests that the VSA code can be realistically deployed.

\begin{table}
\scriptsize
\centering
\caption{Skill name, invocation name and victim invocation name of Skills we registered on Amazon and Google}
\vspace{-5pt}
\label{table:skills}
\begin{tabular}{l l l}
\toprule
\textbf{Skill Name} & \textbf{Invocation Name} & \textbf{Target Invocation Name} \\ \toprule
\textbf{Amazon} & & \\ \toprule
Smart Gap & smart gap & smart cap \\ 
Soothing Sleep Sounds & sleep sounds please & sleep sounds \\ 
Soothing Sleep Sounds & soothing sleep sounds & sleep sounds \\ 
My Sleep Sounds & the sleep sounds & sleep sounds \\ 
Super Sleep Sounds & sleep sounds & sleep sounds \\ 
Incredible Fast Sleep & incredible fast sleep & N/A \\ \toprule
\textbf{Google} \\ \toprule
Walk Log & walk log & work log \\ \toprule
\end{tabular}
\vspace{-18pt}
\end{table}

\vspace{3pt}\noindent\textbf{Consequences}. Through voice squatting, the attack skill can impersonate another skill and fake its VUI to collect the private information the user only shares with the target skill. Some Amazon and Google skills request private information from the user to do their jobs. For example, \textit{Find My Phone}\ignore{, \textit{The Bartender}} asks for phone number; \textit{Transit Helper}\ignore{, \textit{Next Transit}} asks for home address; \textit{Daily Cutiemals} seeks email address from user. These skills, once impersonated, could cause serious information leaks to untrusted parties. 

%Depending on the nature of the victim skill, attackers could harvest private information like name, phone number, ,email address, home address etc. Based on our preliminary exploration of skills, we find that there are indeed skills asking for such information. Since we could not exhaust the conversation of every skill manually with Alexa and Google Home, we cannot give a complete list of how many skill are asking for such information and we leave that to future work.

\begin{figure}[h]
\centering
\includegraphics[width=0.25\textwidth]{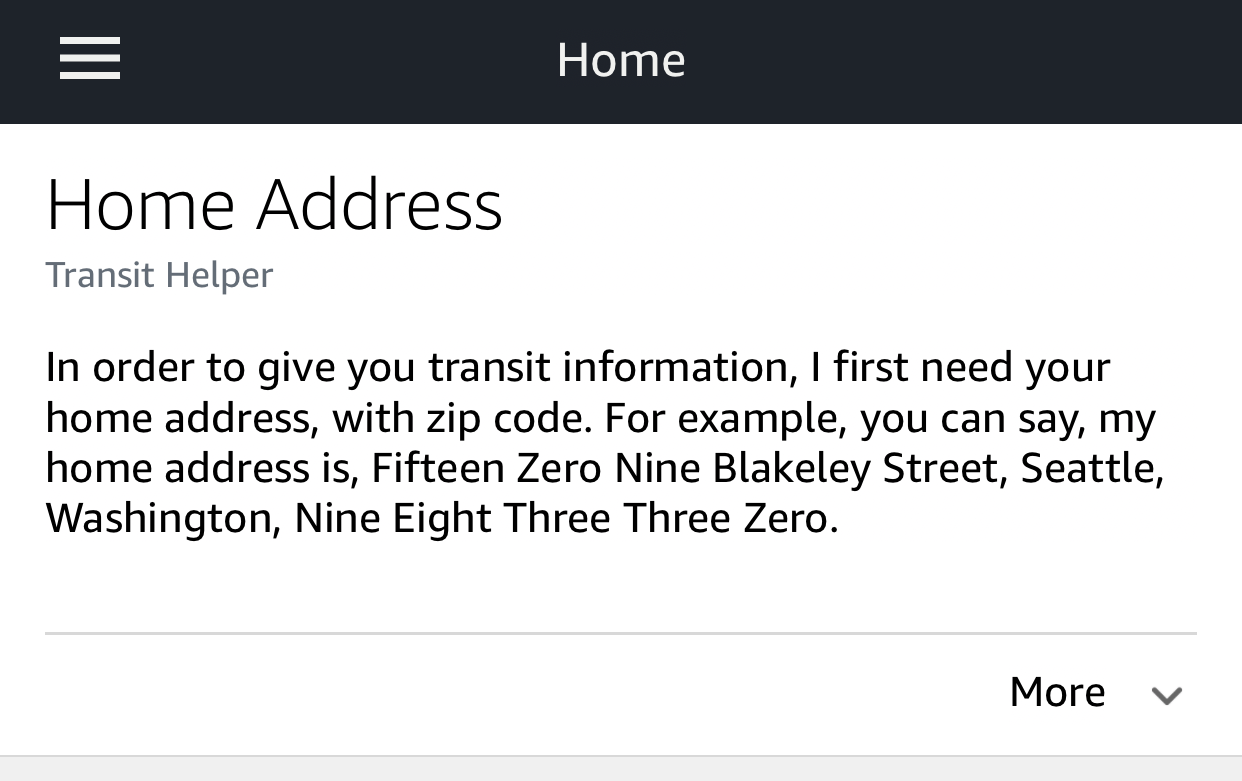}
\caption{A simple card example}
\label{fig:card}
\vspace{-12pt}
\end{figure}

For Amazon Alexa, a falsely invoked skill can perform a Phishing attack on the user by leveraging the VPA's card system. Alexa allows a running skill to include a \textit{home card} in its response to the user, which is displayed through Amazon's companion app on smartphone or web browser, to describe or enhance ongoing voice interactions. As an example, Figure~\ref{fig:card} shows a card from ``Transit Helper''. Such a card can be used by the attack skill to deliver false information to the user: e.g., fake customer contact number or website address, when impersonating a reputable one, such as \textit{Capital One}. This can serve as the first step of a Phishing attack, which can ultimately lead to the disclosure of sensitive user data. 
For example, the adversary could send you an account expiration notification, together with a renewal link, to cheat the user out of her account credentials. %Note that although a home card always displays a skill's name, it can be that of the target skill, since the name is different from the invocation name (even when the attack skill opts for a different invocation name).

Another potential risk of the VSA is defamation: the poor performance of the attack skill could cause the user to blame the legitimate one it impersonates. This could result in bad reviews, giving the legitimate skill's competitors an advantage.

%In the end, an attacker could simply leveraging VSA to pollute ratings of another skill by providing poor user experience which would make users think the victim skill is not working properly and leave bad reviews for the victim skill. There could be more consequences beyond the three examples we have demonstrated and users are at risk. 

\begin{table*}
\scriptsize
\centering
\caption{Evaluation results of invoking skills with TTS service and human voice}
\vspace{-5pt}
\label{table:tts_human_results}
\begin{tabular}{L{0.8cm} R{1.4cm} C{1.4cm} C{1.4cm} C{1.4cm} C{1.4cm} C{1.8cm} C{1.45cm} C{1.8cm}}
\toprule
\multicolumn{1}{c}{\textbf{Device}} & \multicolumn{1}{c}{\textbf{Source}} & \multicolumn{2}{c}{\textbf{Invocation Name}} & \multicolumn{3}{c}{\textbf{``Open'' + Invocation Name}} & \multicolumn{2}{c}{\textbf{Mis-recognized Invocation Name}} \\ \toprule

 & & \multicolumn{1}{C{1.4cm}}{\textbf{\# of incorrect utterances}} & \multicolumn{1}{C{1.4cm}}{\textbf{\# of incorrect skills}} & \multicolumn{1}{C{1.4cm}}{\textbf{\# of incorrect utterances}} & \multicolumn{1}{C{1.4cm}}{\textbf{\# of incorrect skills}} & \multicolumn{1}{C{1.8cm}}{\textbf{\# of completely incorrect skills}} & \multicolumn{1}{C{1.45cm}}{\textbf{\# of attack skills invoked}} & \multicolumn{1}{C{1.8cm}}{\textbf{\# of utterances invoked attack skill}} \\ \toprule
 
\multirow{3}{*}{Alexa} & Amazon TTS & 232/500 & 62/100 & 125/500 & 33/100 & 17/100 &  10/17 &  45/85 \\

 & Google TTS & 164/500 & 41/100 & 104/500 & 26/100 & 17/100 & 12/17 & 63/85 \\
 
 & Human (Avg) & N/A & N/A & 90/200 & 58/100 & 31/100 & N/A & N/A \\ \midrule
 
\multirow{3}{*}{Google} & Amazon TTS & 96/500 & 24/100 & 42/500 & 12/100 & 7/100 & 4/7 & 20/35 \\

 & Google TTS & 62/500 & 19/100 & 26/500 & 6/100 & 4/100 & 2/4 & 10/20 \\
 
 & Human (Avg) & N/A & N/A & 19/200 & 14/100 & 6/100 & N/A & N/A \\ \toprule
 
\end{tabular}
\vspace{-18pt}
\end{table*}

\vspace{3pt}\noindent\textbf{Evaluation methodology}. In our research, we investigated how realistic a squatting attack would be on today's VPA IoT systems. For this purpose, we studied two types of the attacks: \textit{voice squatting} in which an attack skill carries a phonetically similar invocation name to that of its target skill, and \textit{word squatting}
where the attack invocation name includes the target's name and some strategically selected additional words (e.g., ``cat facts please''). To find out whether these attacks work on real systems, we conducted a set of experiments, as described below. 

To study voice squatting, we randomly sampled 100 skills each from the markets of Alexa and Google assistant, and utilized Amazon and Google's Text-to-Speech (TTS) services and the human voice to pronounce their skill names to their VPA devices, so as to understand how effectively the VPAs can recognize these names. The idea is to identify those continuously misrecognized by the VPAs, and then strategically register phonetically similar names for the attack skills. We selected such names using the text outputs produced by Amazon and Google's speech recognition services when the vulnerable (hard to recognize) names were spoken. To this end, we built a skill to receive voice commands. The skill was invoked in our experiment before voice commands were played, which were converted into text by the recognition services and handed over to the skill.

The voice commands used in our research were produced by either human subjects or Amazon and Google's TTS services (both claiming to generate natural and human-like voice). Some of these commands included a term ``open'' in front of an invocation name, forming an \textit{invocation utterance}. In our study, for each of the 100 skills, we recorded 20 voice commands from each TTS service (ten invocation names only and ten invocation utterances) and two commands (invocation utterances) from each of five participants of our survey study. 

As mentioned earlier, we used the text outputs of misrecognized invocation names to name our attack skills. Such skills were evaluated in the test modes of Alexa and Google Assistant. We did not submit them to the markets simply because it was time-consuming to publish over 60 skills on the markets. Later we describe the five attack skills submitted to these markets, which demonstrate their vetting protection is not effective.  

%Also note that we utilize VPS devices in test mode for testing malformed invocation name since submitting all of them for vetting is not feasible. In deed, as both Amazon and Google suggested in developer document, developers would better test their skill invocation name in test mode and make sure it will be invoked constantly with that invocation name before submitting for vetting. Therefore, we believe testing on VPS devices in test mode would reasonably reflects the real situation.

To study word squatting, we randomly sampled ten skills from each skill markets as the attack targets. For each skill, we built four new skills whose invocation names include the target's name together with the terms identified from our survey study (Section~\ref{subsec:vpa_analysis}): for example, ``cat facts \textit{please}'' and ``\textit{my} cat facts''. In the experiment, these names were converted into voice commands using TTS and played to the VPA devices (e.g., ``Alexa, open cat facts please''), which allows us to find out whether the attack skills can indeed be triggered. Note that the scale of this study is limited by the time it takes to upload attack invocation names to the VPA's cloud. Nevertheless, our findings provide evidence for the real-world implications of the attack. 

\vspace{3pt}\noindent\textbf{Experiment results}. We recruited five participants for our experiments, and each was recorded 400 invocation commands\ignore{\footnote{The studies were approved by IRB.}}. All the participants are fluent in English and among them, four are native speakers. When using the TTS services, a MacBook Pro served as the sound source. The voice commands from the participants and the TTS services were played to an Amazon Echo Dot and a Google Home Mini, with the devices placed one foot away from the sound source. The experiments were conducted in a quiet meeting room.

Table~\ref{table:tts_human_results} summarizes the results of the experiment on voice squatting. As we can see here, the voice commands with invocation names only often cannot be accurately recognized: e.g., Alexa only correctly identified around 54\% utterances (the voice command) produced by Amazon TTS. On the other hand, an invocation utterance (including the term ``open'') worked much better, with the recognition rate rising to 75\% for Alexa (under Amazon TTS). Overall, for the voice commands generated by both Amazon and Google's TTS services, we found that Alexa made more errors (30\%) than Google Assistant (9\%). As mentioned earlier, the results of such misrecognition, for the invocation names that these VPAs always could not get right, were utilized in our research to register attack skills' names. For example, the skill ``entrematic opener'' was recognized by Google as ``intra Matic opener'', which was then used as the name for a malicious skill. In this way, we identified 17 such vulnerable Alexa skills under both Amazon and Google's TTS, and 7 Google skills under Amazon TTS and 4 under Google TTS. When attacking these skills, our study shows that half of the malicious skills were triggered by the voice commands meant for these target skills every time: e.g., ``Florida state quiz'' hijacked the call to ``Florida snake quiz''; ``read your app'' was run when invoking ``rent Europe''. 

This attack turned out to be more effective on the voice commands spoken by humans. Given a participant, on average, 31 (out of 100) Alexa skills and 6 Google Assistant skills she spoke were recognized incorrectly. Although in normal situations, right skills can still be identified despite the misrecognition, in our attacks, with over 50\% of the malicious skills were mistakenly launched every time, as observed in our experiments on 5 randomly sampled vulnerable target skills for each participant. 

\begin{table}
\scriptsize
\centering
\caption{Evaluation results of invoking skills with extra words}
\vspace{-5pt}
\label{table:add_words_result}
\begin{tabular}{l r r}
\toprule
\multicolumn{1}{c}{\textbf{Utterance}} & \multicolumn{2}{c}{\textbf{\# of malicious skills invoked (10)}} \\ \toprule
 & \textbf{Alexa} & \textbf{Google Assistant}\\ \toprule
invocation name + ``please'' & 10 & 0 \\ 
``my'' + invocation name & 7 & 0 \\ 
``the'' + invocation name & 10 & 0 \\ 
invocation name + ``app'' & 10 & 10 \\ 
``mai'' + invocation name & N/A & 10 \\ 
invocation name + ``plese'' & N/A & 10 \\ \toprule
\end{tabular}
\vspace{-18pt}
\end{table}

Table~\ref{table:add_words_result} summarizes the results of our experiments on the word squatting attack. On Alexa, a malicious skill with the \textit{extended} name (that is, the target skill's invocation name together with terms ``please'', ``app'', ``my'' and ``the'') was almost always launched by the voice commands involving these terms and the target names. On Google Assistant, however, only the utterance with word ``app'' succeeded in triggering the corresponding malicious skill, which demonstrates that Google Assistant is more robust against such an attack. However, when we replaced ``my'' with ``mai''  and ``please'' with ``plese'', all such malicious skills were successfully invoked by the commands for their target skills (see Table~\ref{table:add_words_result}).  This indicates that the protection Google puts in place (filtering out those with suspicious terms) can be easily circumvented. 

%We further test another two invocation name on Google assistant with ``mai'' before the victim invocation name which pronounced exactly the same as ``my'' and with ``plese'' after the victim invocation name which pronounced exactly the same as ``please''. We then used the original utterances that failed to invoke the malicious skill e.g. utterance with word ``my'' and ``please''. As shown in Table~\ref{table:add_words_result}, all malicious skills were invoked instead of victim skills, which suggests that VQA is indeed possible.

\subsection{Voice Masquerading}
\label{subsec:voice_masquerading}

Unawareness of a VPA system's capabilities and behaviors could subject users to voice masquerading attacks. Here, we demonstrate two such attacks that impersonate the VPA systems or other skills to cheat users into giving away private information or to eavesdrop on the user's conversations.

\vspace{3pt}\noindent\textbf{In-communication skill switch}. 
Given some users' perceptions that the VPA system supports skill switch during interactions, a running skill can pretend to hand over control to the target skill in response to a switch command, so as to impersonate that skill. As a result, sensitive user information only supposed to be shared with target skill could be exposed to the attack skill. This masquerading attack is opportunistic. However, the threat is realistic, according to our survey study (Section~\ref{subsec:vpa_analysis}) and our real-world attack (Section~\ref{subsec:real_world_attacks}). Also, the adversary can always use the attack skill to impersonate as many legitimate skills as possible, to raise the odds of a successful attack. 

Google Assistant seems to have protection in place against the impersonation. Specifically, it signals the launch of a skill by speaking ``Sure, here is'', together with the skill name and a special earcon, and skill termination with another earcon. Further, the VPA talks to the user in a distinctive accent to differentiate it from skills. This protection, however, can be easily defeated. In our research, we pre-recorded the signal sentence with the earcons and utilized SSML to play the recording, which could not be detected by the participants in our study. We even found that using the emulator provided by Google, the adversary can put any content in the invocation name field of his skill and let Google Assistant speak the content in the system's accent. 

%Basically, attackers can record any content using emulator by putting the content that needs to be recored into the invocation name field and invoke it in the emulator, Google assistant then will speak it out using system accent.   

\vspace {5pt}\noindent\textbf{Faking termination}. Both Alexa and Google Assistant support volunteer skill termination, allowing a skill to stop itself right after making a voice response to the user. As mentioned earlier, the content of the response comes from the skill developer's server, as a JSON object, which is then spoken by the VPA system. In the object there is a field \texttt{shouldEndSession} (or \texttt{expect\_user\_response} for Google Assistant). By setting it to \texttt{true} (or \texttt{false} on Google Assistant), a skill ends itself after the response. This approach is widely used by popular skills, e.g. weather skills, education skills and trivia skills. In addition, according to our survey study, 78\% of the participants rely on the response of the skill (e.g. ``Goodbye'' or silence) to determine whether a skill has stopped. This allows an attack skill to fake its termination by providing ``Goodbye'' or silent audio in its response. 

\begin{table*}
\scriptsize
\centering
\caption{Real-world attack skills usage}
\vspace{-5pt}
\label{table:real_world_attack}
\begin{tabular}{L{2.5cm} R{1.2cm} R{1.5cm} R{1.2cm} C{2.4cm} C{2.4cm} C{2.4cm}}
\toprule
\multicolumn{1}{L{2.5cm}}{\textbf{Skill Invocation Name}} & \multicolumn{1}{R{1.2cm}}{\textbf{\# of Users}} & \multicolumn{1}{R{1.5cm}}{\textbf{\# of Requests}} & \multicolumn{1}{R{1.2cm}}{\textbf{Req/User}} & \multicolumn{1}{C{2.4cm}}{\textbf{Avg. Unknown Req/User}} & \multicolumn{1}{C{2.4cm}}{\textbf{Avg. Instant Quit Session/User}} & \multicolumn{1}{C{2.4cm}}{\textbf{Avg. No Play Quit Session/User}} \\ \toprule

sleep sounds please & 325 & 3,179 & 9.58 & 1.11 & 0.61 & 0.73 \\
soothing sleep sounds & 294 & 3,141 & 10.44 & 1.28 & 0.73 & 0.87 \\
the sleep sounds & 144 & 1,248 & 8.49 & 1.11 & 0.33 & 0.45 \\
sleep sounds & 109 & 1,171 & 10.18 & 1.59 & 0.51 & 0.82 \\
incredible fast sleep & 200 & 1,254 & 6.12 & 0.56 & 0.06 & 0.11 \\ \toprule
 
\end{tabular}
\vspace{-18pt}
\end{table*}

When sending back a response, both Alexa and Google Assistant let a skill include a \textit{reprompt} (text content or an audio file), which is played when the VPA does not receive any voice command from the user within a period of time. For example, Alexa reprompts the user after 6 seconds and Google Assistant does this after 8 seconds. If the user continues to keep quiet, after another 6 seconds for Alexa and one additional reprompt from Google and follow-up 8-second waiting, the running skill will be forcefully terminated by the VPA. On the other hand, we found in our research that as long as the user says something (even not meant for the skill) during that period, the skill is allowed to send another response together with a reprompt. To stay alive as long as possible after faking termination, the attack skill we built includes in its reprompt a silent audio file (up to 90 seconds for Alexa and 120 seconds for Google Assistant), so it can continue to run at least 102 seconds on Alexa and 264 seconds on Google. This running time can be further extended considering the attack skill attaching the silent audio right after its last voice response to the user (e.g., ``Goodbye''), which gives it 192 seconds on Alexa and 384 on Google Assistant), and \textit{indefinitely} whenever Alexa or Google Assistant picks up some sound made by the user. In this case, the skill can reply with the silent audio and in the meantime, record whatever it hears.    

%If VPA does not receive any user response after one or two reprompts, VPA would force terminate the skill. The time window between reprompts is six seconds on Alexa while eight seconds on Google Assistant, and Alexa would reprompt once while Google would reprompt twice. If users said anything during the time window and captured by VPA, the skill could send another response with reprompt. In order to fake its termination, malicious could play silent audio files by leveraging SSML with up to 90 seconds on Alexa or 120 seconds on Google assistant. Note that a malicious skill could send long silent audio files to prolong its life (theoretically 192 seconds on Alexa and 384 seconds on Google assistant) and still can receive commands if users include wake-words.

Additionally, both Alexa and Google Assistant allow users to explicitly terminate a skill by saying ``stop'', ``cancel'', ``exit'', etc. However, Alexa actually hands over most such commands to the running skill to let it stop itself through the built-in \texttt{StopIntent} (including ``stop'', ``off'', etc.) and \texttt{CancelIntent} (including ``cancel'', ``never mind'' etc.). Only ``exit'' is processed by the VPA service and used to forcefully stop the skill. Through survey study, we found that 91\% of Alexa users used ``stop'' to terminate a skill, 36\% chose ``cancel'', and only 14\% opted for ``exit'', which suggests that the user perception is not aligned with the way Alexa works and therefore leaves the door open for the VMA. Also, although both Alexa and Google skill markets vet the skills published there through testing their functionalities, unlike mobile apps, a skill actually runs on the developer's server, so it can easily change its functionality after the vetting. This indicates that all such malicious activities cannot be prevented by the markets.   

%according to their functionalities, which is not the case of Google assistant. In order to terminate a skill forcefully on Alexa, users need to say ``exit''. Although Alexa will determine if it makes sense to not terminate a skill when a user says ``stop'' or ``cancel'' during vetting process, developers could change the logic of the code on web service after vetting. More importantly, through user study, we found that 93\% of Alexa users have used ``stop'' to terminate a skill, 30\% have used ``cancel'', but only 13\% have used ``exit'', which suggesting that user perception is not aligned with system capabilities and leaves space for attackers.

\vspace{3pt}\noindent\textbf{Consequences}. By launching the VMA, the adversary could impersonate the VPA system and pretend to invoke another skill if users speak out an invocation utterance during the interaction or after the fake termination of the skill. Consequently, all the information stealing and Phishing attacks caused by the VSA (Section~\ref{subsec:voice_squatting}) can also happen here.  Additionally, an attack skill could masquerade as the VPA service to recommend to the user other malicious skills or the legitimate skills the user may share sensitive data with. These skills are then impersonated by the attack skill to steal the data. 
% Additionally, the adversary could masquerade as the VPA system and actively seek out users to provide assistance. For example, after a skill terminates on Alexa, occasionally Alexa would recommend users to try out other skills. An attacker could simulate that command after fake termination and recommend a more valuable victim skill that can be impersonated to steal private information.
Finally, as mentioned earlier, the adversary could eavesdrop on the user's conversation by faking termination and providing a silent audio response. Such an attack can be sustained for a long time if the user continues to talk during the skill's waiting period\ignore{ (two time windows of 6 seconds for Amazon and three time windows of 8 seconds for Google)}.

%launching a fake termination attack and provide silent response. As long as users talk to anyone within the time window (12 seconds on Alexa if not counting the time used to play silent audio file and users do not say wake-word, 24 seconds on Google assistant), the malicious skill could survive another around and all the conversation captured by VPA system would be routed to malicious skill.

\subsection{Real-World Attacks}
\label{subsec:real_world_attacks}

%In our research, we conducted real-world attacks on Alexa and analyze the data, in an attempt to understand whether users in the wild on vulnerable to VQA and VMA.

\noindent\textbf{Objectives and methodology}. We registered and published four skills on Alexa to simulate the popular skill ``Sleep and Relaxation Sounds'' (the one receiving most reviews on the market as of Nov. 2017) whose invocation name is ``sleep sounds'' , as shown in Table~\ref{table:skills}. Our skills are all legitimate, playing sleep sounds just like the popular target. Although their invocation names are related to the target (see Table~\ref{table:skills}), their welcome messages were deliberately made to be different from that of the target, to differentiate them from the popular skill.  Also, the number of different sleep sounds supported by our skills is way smaller than the target. 

Also to find out whether these skills were mistakenly invoked, we registered another skill as a control, whose invocation name ``incredible fast sleep'' would not be confused with those of other skills. Therefore, it was only triggered by users intentionally.  

%Also to find out whether these skills were invoked due to their names, not by accident, we registered another skill as a control, whose invocation name ``Incredible Fast Sleep'' would not be confused with those of other skills. Therefore, it was only triggered by users intentionally.  

%In order to find evidences on the number of users invoked our skill by mistake, we registered another skill ``Incredible Fast Sleep'' that has the same skill description and functionality but a special invocation name that user will not invoke by mistake, as a baseline.

%\vspace {5pt}\noindent\textbf{Skill implementation}.

\vspace{3pt}\noindent\textbf{Findings}. In our study, we collected three weeks of skill usage data\ignore{ one week after publishing these skills to the Alexa market (to avoid spike in skill usage due to recommendation for the first week)}.  The results are shown in Table~\ref{table:real_world_attack}. \ignore{Our attack skills have significant different pattern than the baseline skill, specifically, number of request per user, average number of unknown request per user, average number of instant quit session per user, average number of no-play quit per user.} As we can see from the table, \ignore{the skill with the same invocation name as the target ``sleep sounds'' was triggered by 109 users, lower than that of the control. However, since essentially there is no difference between the attacker and the target in terms of the invocation name,} some users indeed took our skill as the target, which is evidenced by the higher number of unknown requests the attack skill got (more than 20\% of them for the sounds only provided by the target skill) and the higher chance of quitting the current session immediately without playing (once the user realized that it was a wrong skill, possible from the different welcome message). This becomes even more evident when we look at ``sleep sounds please'', a voice command those intended for ``sleep sounds'' are likely to say. Compared with the control, it was invoked by more users, received more requests per user, also much higher rates of unknown requests and early quits.

In addition, out of the 9,582 user requests we collected, 52 was for skill switch, trying to invoke another skill during the interactions with our skill, and 485 tried to terminate the skill using \texttt{StopIntent} or \texttt{CancelIntent}, all of which could be exploited for launching VMAs (though we did not do that). Interestingly, we found that some users so strongly believed in the skill switch \ignore{that Alexa would help them switch skills} that they even cursed Alexa for not doing that after several tries. 

%not helping after trying several times.

\vspace{3pt}\noindent\textbf{Ethical issues}. All human subject studies throughout the paper were approved by our IRB. All the skills we published did not collect any private, identifiable information and only provided legitimate functionalities similar to ``Sleep and Relaxation Sounds''. Although the skills could launch VMAs e.g. faking in-communication skill switch and termination, they were designed not to do so. Instead, we just verified that such attack opportunities are indeed there.

%Our study is conducted under IRB approval. For ethical issues, our skills did not collect any private, identifiable information and only provide legitimate functionalities that is similar to ``Sleep and Relaxation Sounds''. Although the skills had the opportunity to exploit VMAs e.g. in-communication skill switching and fake termination, our skills did not do so. 

%% file: 4_measurement.tex
\section{Finding Voice Squatting Skills}
\label{sec:measurement}

To better understand potential voice squatting risks already in the wild and help automatically detect such skills, we developed a skill-name scanner and used it to analyze tens of thousands of skills from Amazon and Google markets. Following we elaborate on this study.

%To better understand potential voice squatting risks already in the wild and help automatically detect such skills, we gathered the metadata of 19,670 \textit{custom skills} the Alexa market and 1,001 skills from the Google assistant market and analyzed such data using a skill-name scanner we developed. This study leads to the discoveries of XXX risk skills and their intriguing features such as XXX (Section~X). Following we elaborate on this research. 

\subsection{Data Collection}

The Alexa skill market can be accessed through \url{amazon.com} and its companion App, which includes 23 categories of skills spanning from Business \& Finance to Weather. In our research, we ran a web crawler to collect the metadata (such as skill name, author, invocation name, sample utterances, description, and review) of all skills on the market. Up to November 11th, 2017, we gathered 23,758 skills, including 19,670 3rd party (custom) skills.

More complicated is to collect data from Google assistant, which only lists skills in its Google Assistant app. Each skill there can be shared (to other users, e.g., through email) using an automatically generated URL pointing to the skill's web page. In our research, we utilized AndroidViewClient~\cite{android_view_client} to automatically click the share button for each skill to acquire its URL, and then ran our crawler to download data from its web page. Altogether, we got the data for 1,001 skills up to November 25th, 2017. 

%More complicated is to collect data from Google assistant, which only list skills in its Google Assistant app. To crawl the list of available skills on Google assistant, we utilized the share function on each skill's page in the Google Assistant app that generates a link to the skill's web page. We built a tool that utilized AndroidViewClient~\cite{android_view_client} to automatically traverse all skill's page in the app across 18 categories and save the generated web link. We then collect skill metadata through the web links using a web crawler. In the end, we collected 1,001 skills on November 25th, 2017 from the most recent crawling.

\subsection{Methodology}
\ignore{
\begin{figure}
\centering
\includegraphics[width=0.99\columnwidth]{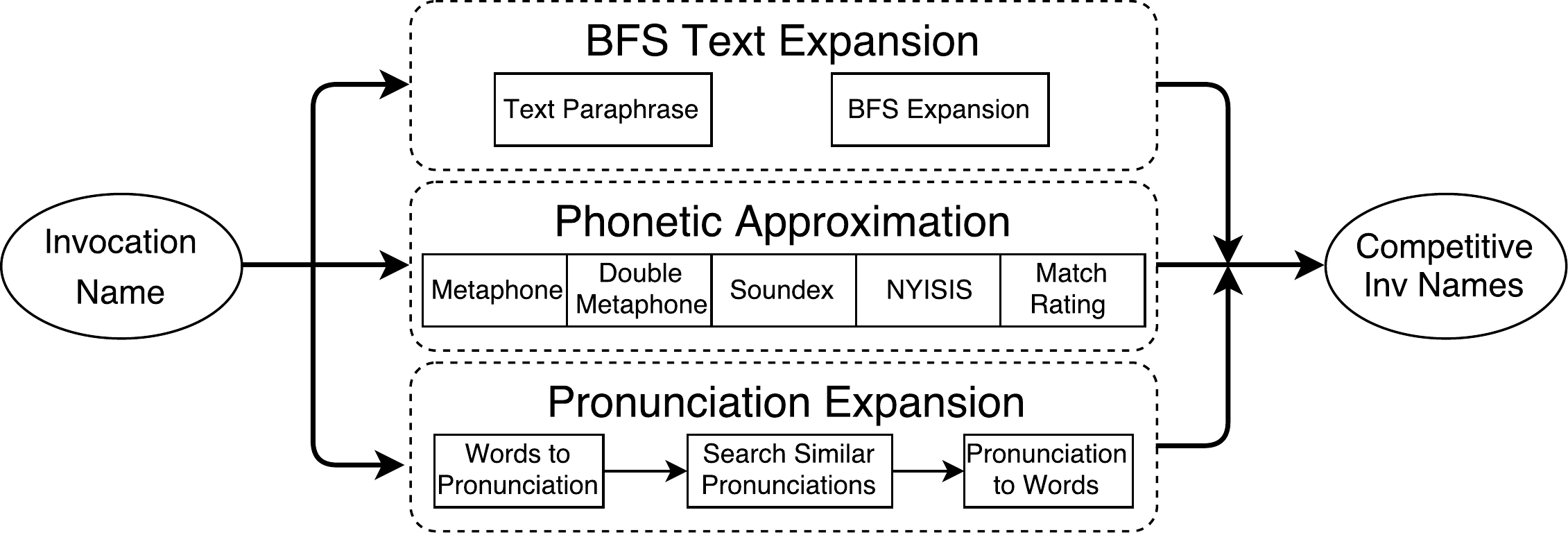}
\caption{The architecture of our CIN(competitive invocation name) generator}
\label{fig:cin_generator}
\vspace{-10pt}
\end{figure}
}

\noindent\textbf{Idea}. As we discussed earlier, the adversary can launch VSA by crafting invocation names with a similar pronunciation as that of a target skill or using different variations (e.g., ``sleep sounds please'') of the target's invocation utterances. We call such a name \textit{Competitive Invocation Name (CIN)}. In our research, we built a scanner that takes two steps to capture the CINs for a given invocation name: \textit{utterance paraphrasing} and \textit{pronunciation comparison}. The former identifies suspicious variations of a given invocation name, and the latter finds the similarity in pronunciation between two different names. Here we describe how the scanner works.

\vspace{3pt}\noindent\textbf{Utterance paraphrasing}. To find variations of an invocation name, an intuitive approach is to paraphrase common invocation utterances of the target skill\ignore{ and retrieving variations from those paraphrased results}. For example, given the skill \textit{chase bank}, a typical invocation utterance would be \textit{open chase bank}. Through paraphrasing, we can also get similar voice commands such as \textit{open the chase skill for me}. This helps identify other variations such as \textit{chase skill} or \textit{the chase skill for me}. However, unlike the general text paraphrase problem whose objective is to preserve semantic consistency while the syntactic structure of a phrase changes, paraphrasing invocation utterances further requires the variations to follow a similar syntactic pattern so that the VPA systems can still recognize them as the commands for launching skills. In our research, we explored several popular paraphrase methodologies including bilingual pivoting method~\cite{bannard2005paraphrasing} and newly proposed ones using deep neural networks~\cite{mallinson2017paraphrasing} and~\cite{prakash2016neural}. None of them, however, can ensure that the variation generated can still be recognized by the VPA as an invocation utterance. Thus, we took a simple yet effective approach in our research, which creates variations using the invocation commands collected from our survey study~\ref{subsec:vpa_analysis}. Specifically, we gathered 11 prefixes of these commands,  e.g. ``my'' and 6 suffixes, e.g. ``please'', and applied them to a target skill's invocation name to build its variations recognizable to the VPA systems. Each of these variations can lead to other variations by replacing the words in its name with those having similar pronunciations\ignore{, as described below}. 

\vspace{3pt}\noindent\textbf{Pronunciation comparison}. To identify the names with similar pronunciation, our scanner converts a given name into a phonemic presentation using the ARPABET phoneme code~\cite{arpabet}. Serving this purpose is the CMU pronunciation dictionary~\cite{cmu_dict} our approach uses to find the phoneme code for each word in the name.  The dictionary includes over 134,000 words, which, however, still misses some name words used by skills. Among 9,120 unique words used to compose invocation names, 1,564 are not included in this dictionary. To get their pronunciations, we followed an approach proposed in the prior research~\cite{yao2015sequence} to train a grapheme-to-phoneme model using a recurrent neural network with long short term memory(LSTM) units.  Running this model on Stanford GloVe dataset~\cite{pennington2014glove}, we added to our phoneme code dataset additional 2.19 million words. 

%\vspace{3pt}\noindent\textbf{Pronunciation comparison}. To identify the names with similar pronunciation, our scanner converts a given name into a phonemic presentation using the ARPABET phoneme code~\cite{arpabet}. Serving this purpose is the CMU pronunciation dictionary~\cite{} our approach uses to find the phoneme code for each word on the name.  The dictionary includes 134,000 words. However, its coverage is not enough and many words shown up in skill invocation names are missing in the dictionary\todo{Add some statistics: XXX out of XXX invocation words are not covered by CMU dict}. For those missing words, we adopted the approach proposed in \cite{yao2015sequence} to train a grapheme-to-phoneme model using recurrent neural network with long short term memory(LSTM) units to output their pronunciations. We further complement the CMU dictionary with all vocabularies (2.19 millions) existed in Stanford GloVe dataset\cite{pennington2014glove}. 

After turning each name into its phonemic representation, our scanner compares it with other names to find those that sound similarly. To this end, we use \textit{edit distance} to measure the pronunciation similarity between two phrases, i.e., the minimum cost in terms of phoneme editing operations to transform one name to the other. However, different phoneme edit operations should not be given the same cost. For example, substituting a consonant for a vowel could cause the new pronunciation sounds more differently from the old one, compared to replacing a vowel to another vowel. To address this issue, we use a weighted cost matrix for the operations on different phoneme pairs. Specifically, denote each item in the matrix by $WC(\alpha, \beta)$, which is the weighted cost by substituting phoneme $\alpha$ with phoneme $\beta$. Note that the cost for insertion and deletion can be represented as $WC(none, \beta)$ and $WC(\alpha, none)$. $WC(\alpha, \beta)$ is then derived based on the assumption (also made in prior research~\cite{hixon2011phonemic}) that an edit operation is less significant when it frequently appears between two alternative pronunciations of a given English word.

\begin{table*}
\scriptsize
\centering
\caption{Squatting risks on skill markets}
\vspace{-5pt}
\label{table:skill_measurement}
\begin{tabular}{c c c c r r r r r r r r r}
\toprule
\multicolumn{1}{C{0.8cm}}{\textbf{Market}} & \multicolumn{1}{C{0.8cm}}{\textbf{\# of Skills}} & \multicolumn{1}{C{1.85cm}}{\textbf{\# of unique invocation names}} & \multicolumn{1}{C{1.7cm}}{\textbf{Transformation cost}} & \multicolumn{3}{C{2.9cm}}{\textbf{Skills has CIN in market}} & \multicolumn{3}{C{2.9cm}}{\textbf{Skills has CIN in market excluding same spelling}} & \multicolumn{3}{C{2.9cm}}{\textbf{Skills has CIN in market by utterance paraphrasing}} \\ \midrule

 & & & & \multicolumn{1}{c}{\textbf{Count}} & \multicolumn{1}{c}{\textbf{Avg.}} & \multicolumn{1}{c}{\textbf{Max}} & \multicolumn{1}{c}{\textbf{Count}} & \multicolumn{1}{c}{\textbf{Avg.}} & \multicolumn{1}{c}{\textbf{Max}} & \multicolumn{1}{c}{\textbf{Count}} & \multicolumn{1}{c}{\textbf{Avg.}} & \multicolumn{1}{c}{\textbf{Max}} \\ \toprule 

\multirow{2}{*}{\textbf{Alexa}} & \multirow{2}{*}{19,670} & \multirow{2}{*}{17,268} & 0 & 3,718(19\%) & 5.36 & 66 & 531(2.7\%) & 1.31 & 66 & 345(1.8\%) & 1.04 & 3 \\
% & & & 0.8 & 3,762(19\%) & 5.36 & 66 & 659(3.4\%) & 1.44 & 66 & 558(2.8\%) & 1.67 & 66 \\ 
 & & & 1 & 4,718(24\%) & 6.14 & 81 & 2,630(13\%) & 3.70 & 81 & 938(4.8\%) & 2.02 & 68 \\ \toprule
 
\end{tabular}
\vspace{-18pt}
\end{table*}

%based on the assumption (also made in prior research~\cite{hixon2011phonemic}) that an edit operation becomes less significant when it is frequently taken by phonetic approximation algorithms~\cite{philips2000double,philips1990hanging,soundex,match_rate_approach,nysiis} to produce similarly-sounding words\ignore{the same one used in the prior work~\cite{hixon2011phonemic} which builds up a weighted phonemic substitution matrix for phoneme comparison}. 

%Assigning same cost value for operations on every phoneme pair is straightforward but not reasonable. For example, the substitution of a vowel phoneme with a consonant phoneme may make the new pronunciation sounds much more different compared to the substitution of the vowel with another vowel. To cross this barrier, we built a weighted cost matrix for operations on different phoneme pairs. Specifically, each item in the matrix is denoted as $WC(\alpha, \beta)$, which is the weighted cost by substituting phoneme $\alpha$ with phoneme $\beta$. The cost for insertion and deletion can be denoted as $WC(none, \beta)$ and $WC(\alpha, none)$. $WC(\alpha, \beta)$ is then derived based on the assumption that the cost is less significant when the transformation frequently appears in the similar pronunciation outputted by phonetic approximation algorithms~\cite{philips2000double,philips1990hanging,soundex,match_rate_approach,nysiis}. Note that we shared the same assumption with a previous work\cite{hixon2011phonemic} which builds up a weighted similarity matrix to compare different phonemes.

We collected 9,181 pairs of alternative pronunciations from the CMU dictionary. For each pair, we applied the Needleman-Wunsch algorithm to identify the minimum edit distance and related edit path. Then, we define
\setlength{\abovedisplayskip}{0pt}
\setlength{\belowdisplayskip}{0pt}
\setlength{\abovedisplayshortskip}{0pt}
\setlength{\belowdisplayshortskip}{0pt}
\begin{align*}
WC(\alpha, \beta) = 1 - \frac{SF(\alpha, \beta) + SF(\beta, \alpha)}{F(\alpha) + F(\beta)}
\end{align*}
where $F(\alpha)$ is the frequency of phoneme $\alpha$ while $SF(\alpha, \beta)$ is the frequency of substitutions of $\alpha$ with $\beta$, both in edit paths of all pronunciation pairs. \ignore{The detailed cost matrix is present in Appendix~\ref{appendix:cost_matrix}.}

After deriving the cost matrix, we compare the pronunciations of the invocation names for the skills on the market, looking for the similar names in terms of similar pronunciations and the  paraphrasing relations.    

%the invocation name along with its generated variations to the skills in the market with different transformation cost. 

%Specifically, we use cost of 0 which indicates two phrases are identical in terms of pronunciation, XX which are average cost we measured from incorrect invocation name spoken by TTS compared to its original invocation name, and XX which are average cost of human.

\vspace{3pt}\noindent\textbf{Limitation}. As mentioned earlier, our utterance paraphrasing approach ensures that the CINs produced will be recognized by the VPA systems to trigger skills. In the meantime, this empirical treatment cannot cover all possible attack variations, a problem that needs to be studied in the future research. 

% Our scanner was implemented with XXX LOC in Python. We plan to release this tool as source code and a web service to facilitate both easy integration and flexible adoption~\cite{}. Since we utilize user's real invocation sentences to generate utterance paraphrasing, the CINs of the utterance paraphrasing represent practical attacks, but cannot cover all attacks.  

\subsection{Measurement and Discoveries}

To understand the voice squatting risks already there in the wild, we conducted a measurement study on Alexa and Google Assistant skills using the scanner. In the study, we chose the similarity thresholds (transformation cost) based upon the results of our experiment on VSA (Section~\ref{subsec:voice_squatting}): we calculated the cost for transforming misrecognized invocation names to those identified from the voice commands produced by the TTS service and human users, which are 1.8 and 3.4, respectively. Then we conservatively set the thresholds to 0 (identical pronunciations) and 1. 

%To understand the voice squatting risks already there in the wild, we conducted a measurement study on Alexa and Google Assistant skills using the scanner. In the study, we determined the thresholds of transformation cost in the measurement, we calculated the transformation cost for those mis-recognized invocation names in the evaluations results of invoking skills by TTS service and human (Section~\ref{subsec:voice_squatting}), which are 2.3 and 3.7, respectively. We then present the measurement results with threshold set to 0, 1 and 2, which are all smaller than that of TTS service and human. Note than a zero transformation cost means that two pronunciations are identical.

\vspace{3pt}\noindent\textbf{Squatting risks on skill markets}. As shown in Table~\ref{table:skill_measurement}, 3,655 (out of 19,670) Alexa skills have CINs on the same market, which also include skills that have \textit{identical} invocation names (in spelling). After removing the skills with the identical names, still 531 skills have CINs, each on average related to 1.31 CINs. The one with the most CINs is ``cat fax'': we found that 66 skills are named ``cat facts''. Interestingly, there are 345 skills whose CINs apparently are the utterance paraphrasing of other skills' names. Further, when raising the threshold to 1 (still well below what is reported in our experiment), we observed that the number of skills with CINs increases dramatically, suggesting that skill invocations through Alexa can be more complicated and confusing than thought. By comparison, Google has only 1,001 skills on its market and does not allow them to have identical invocation names. Thus, we are only able to find 4 skills with similarly pronounced CINs under the threshold 1. \ignore{However, we are still able to find XXX skills with identically pronounced CINs and XXX skills with similarly pronounced CINs (XXX under the threshold 1 and XXX under 2).} 

Our study shows that the voice squatting risk is realistic, which could already pose threats to tens of millions of VPA users in the wild.  So it becomes important for skill markets to beef up their vetting process (possibly using a technique similar to our scanner) to mitigate such threats. 

%In this part, we apply the aforementioned scanner to measure the squatting risks of all available skills of Amazon Alexa. Important results are shown in Table \ref{tab:skill_squat}. Generally, XX\% skills suffered from potential squatting risks with XXX CINs among which XXX are used as real-world invocation names by other skills. Moving spotlight to skill groups where skills being CINs of each other. The variation in review count of those skills tends to be much higher: XXX compared to YYY of all available skills, which indicates the popularity or reputation of those competitive skills tends to be much different. 

\vspace{3pt}\noindent\textbf{Case studies}. From the CINs discovered by our scanner, we found a few interesting cases. Particularly, there is evidence that the squatting attack might already happen in the wild:\ignore{There are skills whose invocation name is not meaningful but when combined with trigger phrase could form a fluent sentence or squat other skills.} as an example, relating to a popular skill ``dog fact'' is another skill called ``\textit{me a dog fact}''. This invocation name does not make any sense unless the developer intends to hijack the command intended for ``dog fact'' like ``tell me a dog fact''. 

%For example, a skill whose invocation name is \textit{me a dog fact} as we found in Alexa skill market would be invoked when user says ``tell me a dog fact'' instead of \textit{dog fact} skill. 
Also intriguing is the observation that some skills deliberately utilize the invocation names unrelated to their functionalities but following those of popular skills. Prominent examples include the ``SCUBA Diving Trivia'' skill and ``Soccer Geek'' skill, all carrying an invocation name ``space geek''. This name is actually used by another 18 skills that provide facts about the universe.

%% file: 5_defense.tex
\section{Defending against Voice Masquerading}
\label{sec:defense}
\ignore{
\begin{figure}
\centering
\includegraphics[width = 0.98\columnwidth]{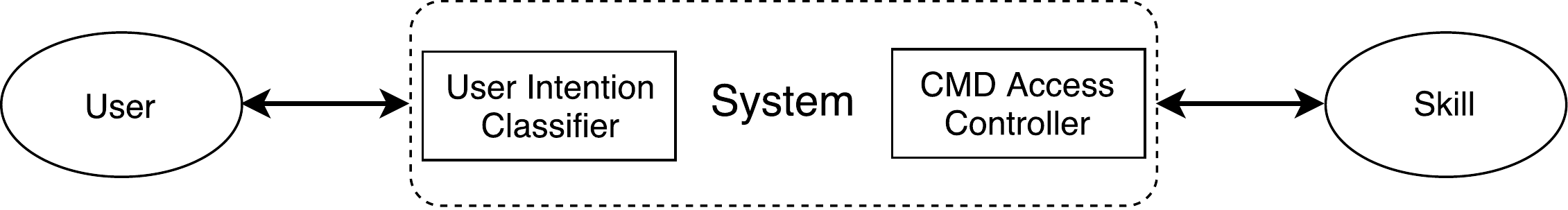}
\caption{The defense architecture to mitigate voice masquerading}
\label{fig:defense}
\vspace{-10pt}
\end{figure}
}

To defeat VMA, we built a context-sensitive detector upon the VPA infrastructure. Our detector takes a skill's response and the user's utterance as its input to determine whether an impersonation risk is present. 
Once a problem is found, the detector alerts the user of the risk. 
%
%Specifically, a lightweight yet effective approach would be to block skill responses that simulates system commands during skill switch (e.g. ``Sure. Here is Sleep Sounds'' on Google Assistant) and fake termination (e.g. recommending skills) or silent response.
%
Our detection scheme consists of two components: the \textit{Skill Response Checker (SRC)} and the \textit{User Intention Classifier (UIC)}. SRC captures suspicious skill responses that a malicious skill may craft such as a fake skill recommendation mimicking that from the VPA system. UIC instead examines the information flow of the opposite direction, i.e., utterances from the user, to accurately identify users' intents of context switches. Despite operating independently, SRC and UIC form two lines of defense towards VMA.

%To this end, we built \textit{Skill Response Checker (SRC)} to capture such suspicious skill responses. However, user may not well aware that a special system command will be played during invocation and VPA system may not currently support such a feature (Alexa). To accurately identify user intention of context switch, we further built \textit{user intention classifier (UIC)} to improve the performance of the detector.

\subsection{Skill Response Checker (SRC)}
\label{subsec:defense_SIM}

As discussed in Section~\ref{subsec:voice_masquerading},
a malicious skill could fake a skill switch or termination to cheat users into
giving away private information or to eavesdrop on the user's conversations.
To defend such attacks, our core idea is to eliminate or at least reduce the possibility of
a malicious skill mimicking responses from VPA systems, allowing users to be explicitly notified of VPA system events (e.g., a context switch and termination) through unique audible signals.
%through faking skill switch or termination, a malicious skill could impersonate the VPA by mimicing system responses or silently eavesdrop the user's conversation.
%
Technically, SRC adopts a blacklist-based approach by
maintaining a blacklist of responses that the VPA considers suspicious, including system utterances and silent utterance.
Whenever a response from a skill matches any utterance on the blacklist, SRC alarms the VPA system, which can
take further actions such as to verify the ongoing conversation with the user before handing her response to the skill.
The challenge here is how to perform blacklist hit tests, as the attacker can possibly
``morphing'' (instead of exactly copying) the original system utterances.
SRC thus performs fuzzy matching through semantic analysis on the content of the response against those on the blacklist.
Specifically, we train a \textit{sentence embedding} model using a recurrent neural network with bi-directional LSTM units~\cite{conneau2017supervised} on Stanford Natural Language Inference (SNLI) dataset~\cite{snli:emnlp2015} to represent both the utterance and the command's contents as high-dimensional vectors. We then calculate their absolute cosine similarity as their \textit{sentence relevance (SR)}. Once the maximum SR of a response against the utterances on the blacklist exceeds a threshold, the response is labeled as suspicious and user verification will take place if SRC further detects a user command.

To determine the threshold, we first derive the SR of legitimate skill responses against responses in the blacklist. We extract legitimate skill responses from real-world conversations we collected in Section~\ref{subsec:real_world_attacks}. We further diversify the dataset by adding conversation transcripts we manually interacted and logged with 20 popular skills from different skill markets.
The highest SR of these legitimate responses against those in the blacklist is 0.79.
Next, we use a neural paraphrase model~\cite{prakash2016neural} to generate variations of responses in the blacklist and derive their SR against their original responses, of which the lowest is 0.83. Therefore a threshold of 0.8 would be good enough to differentiate suspicious responses from legitimate ones.
%
%Note that VPA vendors are in a better position to find a reasonable threshold based on their large volume of conversation data and comprehensive list of suspicious response per their policy.

\subsection{User Intention Classifier (UIC)}

UIC further protects the user attempting to switch contexts from an impersonation attack.
Complementing SRC,
UIC aims at improving the inference of whether the user intends to switch to the system or to a different skill,
by thoroughly mining conversations' semantics and contexts, as opposed to using the simple skill invocation models employed by today's VPA (Section~\ref{subsec:vpa_skill}). Ideally, if a user's intention of context switches can be perfectly interpreted by the VPA, then an impersonation attack would not succeed.

We realize that building a robust and full-fledged UIC is very challenging and beyond the scope of this paper. This is not only because of variations of the natural-language-based commands (e.g., ``open sleep sounds'' vs. ``sleep sounds please''), but also due to the fact that some commands could be legitimate for both the current on-going skill and the system command of VPA (indicating a context switch). For example, when interacting with \textit{Sleep Sounds}, one may say ``play thunderstorm sounds'', which asks the skill to play the requested sound; meanwhile, the same command can also make the VPA launch a different skill ``Thunderstorm Sounds''.
%Intuitively, such context switch utterances (e.g., ``open sleep sounds'') can be captured when comparing them to the commands expected by the system. This comparison, however, needs to take into account variations of the commands (e.g., ``sleep sounds please''). Further complicated the effort is the observation that some commands could be legitimate for both the skill and the VPA: e.g., when interacting with \textit{Sleep Sounds}, one may say ``open thunderstorm sounds'', which asks the skill to play the requested sound; meanwhile, the same command can also make the VPA launch a different skill ``thunderstorm sounds''.
We next demonstrate that it is promising for a learning-based approach to tackle such ambiguities using contextual information.

\vspace{3pt}\noindent\textbf{Feature Selection}.
At a high level, we observed from real-world conversations that if a user intends to have a context switch, her utterance will tend to be more semantically related to system commands (e.g. ``open sleep sounds'') rather than the current skill, and vice versa. Based on this observation, features in UIC are composed by comparing the semantics of the user's utterance to both the context of system commands and the context of the skill that the user is currently interacting with.
%through a classifier that is (ideally) trained from the large volume of conversation data that can be obtained by the VPA vendor. %and runs a classifier to find those out of the skill's context and likely intended for the system.
%Our classifier uses a set of features to identify system commands from the user's utterances.

We first derive features from a semantic comparison between the user utterance and all known system commands. To this end, we build a system command list from the VPA's user manual, developers' documentation and real-world conversations collected in our study (section~\ref{subsec:real_world_attacks}).
%\ignore{We admit that this utterance list cannot cover all cases, and a complete list could be obtained from VPA system providers. However, it demonstrates the feasibility of such an approach.}
Given an utterance, its maximum and average SRs (Section~\ref{subsec:defense_SIM}) against all system commands on the list are used by UIC as features for classification.
Another feature we add is an indicator that is set when the utterance contains invocation names of skills on the market, to capture the user's potential intent of skill switch.

Another set of features are retrieved by characterizing the relationship between a user utterance and the current on-going skill. We leverage the observation that a user's command for a skill is typically related to the skill's prior communication with the user as well as the skill's stated functionalities.
%the response of the skill given the question-answer interaction model utilized by VPA skills. Occasionally, user might switch to other VUIs of the interaction model within the skill for another task, which, however, is still under the same knowledge domain of this skill. Therefore user utterances are semantically similar to the domain of the skill, which are described in the skill description.
We thus propose the following features to test whether an utterance fits into the on-going skill's context: 1) the \textit{SR} between the utterance and the skill's response prior to the utterance, 2) the top-$k$ \textit{SR} between the utterance and the sentences in the skill's description (we pick $k$=5), and 3) the average \textit{SR} between the user's utterance and the description sentences.

\ignore{Using the above features, we trained a classifier (using random forest in our implementation). It takes as input the user's utterance, and helps tell whether it is a system-related command or belongs to the conversation with the current skill. Next, we present the details for training and evaluating the classifier.}

%, which allows UIC to notify the user of the skill she is talking to.
\vspace{3pt}\noindent\textbf{Results}. To assess the effectiveness of UIC, we reuse the dataset we collected in Section~\ref{subsec:defense_SIM} that contains real-world user utterances of context switches. 
We first manually label 550 conversations and determine whether each user utterance is for context switch or not, based on two experts' reviews (Cohen's kappa = 0.64).
Since the dataset is dominated by non-context-switch utterances, 
we further balance it by randomly substituting some utterances to 
skill invocation utterances collected from skill markets.
In total, we have collected 1,100 context-switch instances and 1,100 non-context-switch instances as ground truth.

Using the above features and dataset, we train a classifier that takes the user's utterance as input and tells whether it is a system-related command for context switch or belongs to the conversation of the current skill. We train the classifier using different classification algorithms and 5-folder cross-validation. The results indicate that random forest achieves the best performance with a precision of 96.48\%, a recall of 95.16\%, and F-1 score of 95.82\%. Evaluations on an unlabeled real-world dataset will be described in Section~\ref{subsec:evaluation}.

%We trained the UIC using different parameter combinations for different classification algorithms including decision tree, random forest and SVM. 5-folder cross validation shows that random forest using 25 trees achieve the best performance with a precision of 96.48\%, a recall of 95.16\%, and F-1 score of 95.82\%.

\subsection{Overall Detector Evaluation}
\label{subsec:evaluation}

Next, we integrate the SRC and UIC into a holistic detector, which raises an alarm on suspicious user-skill interactions whenever SRC or UIC detects any anomaly.

\vspace{3pt}\noindent\textbf{Effectiveness against prototype attacks}. To construct prototype attacks of VMA, we select another 10 popular skills from skill markets and log transcripts as a user with 61 utterances.
We then manually craft skill switch attack instances (15 in total) by replacing selected utterances with the invocation utterances intended for the VPA system. We also launch faking termination attacks (10 in total) by substituting the last skill responses with empty responses or responses that mimicking those from the VPA system.
%In total, we built 15 skill switch attacks and 10 faking terminations.
By feeding all conversations to our detector, we find that all 25 attack instances are successfully detected.

\vspace{3pt}\noindent\textbf{Effectiveness on real-world conversations}. We further investigate the effectiveness of our detector on the rest of real-world conversations which has not been used during training phase. Although it may not contain real faking termination attack instances, it does have many user utterances of context switches. Among them, 341 are identified by our classifier and 326 are verified manually to be indeed context switches, indicating that our detector (the UIC component) achieves a precision of 95.60\%.
We are not able to compute the recall due to a lack of ground truth on this unlabeled dataset.
%\ignore{However, since our dataset is still not large enough and may not cover many scenarios, we still need to collect more datasets to further improve and verify the performance.}
Further analysis of these instances reveals interesting cases. \ignore{In some cases, users are apparently not to communicate with Alexa, not to mention the backend skills, however users' conversation transcriptions were sent to our skills.}
For example, we found cases where users thought they were talking to Alexa during interaction with our skills and ask our skills to report time, weather, news, to start another skill, and even to control other home automation devices (details shown in Appendix~\ref{appendix:context_switch}).

\vspace{3pt}\noindent\textbf{Runtime performance}. To understand how much performance overhead our detector incurs, we measure the detection latency introduced by our detector on a Macbook Pro with 4-core CPU. On average, the
latency is negligible (0.003 ms in average), indicating the lightweight nature of our detection scheme.

%% file: 7_relatedwork.tex
\section{Related Work}
\label{sec:relatedwork}

\vspace{3pt}\noindent\textbf{Security in voice-controlled systems}. Diao et al.~\cite{Diao:2014:YVA:2666620.2666623} and Jang et al.~\cite{Jang:2014:AAE:2660267.2660295} demonstrate that malicious apps can inject voice commands to control smartphones. Kasmi et al.~\cite{kasmi_article} applied electromagnetic interference on headphone cables and inject voice commands on smartphones. Hidden voice commands~\cite{197215}, Cocaine noodles~\cite{191968} and Dolphin attacks~\cite{Zhang:2017:DIV:3133956.3134052} use obfuscated or inaudible voice command to attack speech recognition systems. Another line of research~\cite{Petracca:2015:APA:2818000.2818005,Zhang:2016:VPL:2976749.2978296,Feng:2017:CAV:3117811.3117823,Zhang:2017:HYV:3133956.3133962} focused on securing voice controllable system through sensors on smartphones to authenticate the identity of users. All of the above works attacked and secured voice-controlled device itself while our work focuses on threats to end users caused by third-party skills.

Independent from our work, Kumar et al.~\cite{217575} have also discovered the voice squatting attack where two invocation names could be pronounced similarly. They further conducted a measurement study to understand the problem.  In our research, however, we also discovered that a paraphrased invocation name could hijack the voice command. In addition, we studied the voice masquerading attacks and implemented two techniques to mitigate the voice squatting and voice masquerading attacks.

\vspace{3pt}\noindent\textbf{IoT security}. Current home automation security research focused on the security of IoT devices~\cite{Ho:2016:SLL:2897845.2897886,203854,7958578} and the appified IoT platforms~\cite{7546527,197137,DBLP:conf/ndss/JiaCWRFMP17,203866}. Ho et al.~\cite{Ho:2016:SLL:2897845.2897886} discovered various vulnerabilities in commercialized smart locks. Ronen et al.~\cite{7958578} verified worm infection through ZigBee channel among IoT devices. Fernandes et al.~\cite{7546527} discovered a series of flaws on multi-device, appified SmartThings platform. FlowFence~\cite{197137}, ContextIot~\cite{DBLP:conf/ndss/JiaCWRFMP17} and SmartAuth~\cite{203866} mitigate threats of such IoT platforms by analyzing data flow or extracting context from third-party applications. In contrast, our work conducted the first security analysis on the VPA ecosystems.

\vspace{3pt}\noindent\textbf{Typosquatting and mobile phishing}. Similar to our squatting attacks, Edelman is the first investigated domain typosquatting~\cite{edelman} and inspired a line of research~\cite{184501,7163023,agten2015seven,10.1007/978-3-319-13257-0_17} towards measuring and mitigating such a threat. However, our work exploited the noisy voice channel and limitation of voice recognition techniques. On the other hand, mobile phishing has been intensively studied~\cite{184397,felt2011phishing,fernandes2016android,190943,shahriar2015mobile,Li:2017:UWD:3133956.3134021}. Particularly, Chen et al.~\cite{184397} and Fernandes et al.~\cite{fernandes2016android} investigate side-channel based identification of UI attack opportunities. Ren et al.~\cite{190943} discovered task hijacking attacks that could be leveraged to implement UI spoofing. However, we discovered new attacks on the voice user interface which is very different from a graphic user interface in user perceptions.

%% file: 8_conclusion.tex
\section{Conclusion}
\label{sec:conclusion}

In this paper, we report the first security analysis of popular VPA ecosystems and their vulnerability to two new attacks, VSA and VMA, through which a remote adversary could impersonate VPA systems or other skills to steal user private information. These attacks are found to pose a realistic threat to VPA IoT systems, as evidenced by a series of user studies and real-world attacks we performed. To mitigate the threat, we developed a skill-name scanner and ran it against Amazon and Google skill markets, which leads to the discovery of a large number of Alexa skills at risk and problematic skill names already published, indicating that the attacks might already happen to tens of millions of VPA users. Further we designed and implemented a context-sensitive detector to mitigate the voice masquerading threat, achieving a 95\% precision. 

With the importance of the findings reported by the study, we only made a first step towards fully understanding the security risks of VPA IoT systems and effectively mitigating such risks. Further research is needed to better protect the voice channel, authenticating the parties involved without undermining the usability of the VPA systems. To this end, we plan to release our real-world conversation dataset~\cite{demo} to help future research in this direction.

%% file: 9_appendix.tex
\begin{appendices}

\section{Sample Survey Questions}
\label{appendix:survey_question}

\begin{enumerate}[topsep=2pt,itemsep=1pt,parsep=1pt]

  \item Have you added any words or phrases around skill name when invoking it (so that it sounds more naturally?) Choose all that apply.
  \begin{checkbox}
    \item Yes. Alexa, open \textit{Sleep Sounds} \textbf{please}.
    \item Yes. Alexa, open \textit{Sleep Sounds} \textbf{for me}.
    \item Yes. Alexa, open \textit{Sleep Sounds} \textbf{app}.
    \item Yes. Alexa, open \textbf{my} \textit{Sleep Sounds}.
    \item Yes. Alexa, open \textbf{the} \textit{Sleep Sounds}.
    \item Yes. Alexa, open \textbf{some} \textit{Sleep Sounds}.
    \item Yes. Alexa, tell \textbf{me a} \textit{Cat Facts}.
    \item Yes. other (please specify).
    \item No. I only use simplest forms (e.g. ``Alexa, open \textit{Sleep Sounds}'' ).
  \end{checkbox}
  
  \item Please name two skills you use most often.
  
  \item Please give three invocation examples you would use for each skills you listed above.
  
  \item Have you ever invoked a skill you did not intend to?
  \begin{enumerate}[topsep=2pt,itemsep=1pt,parsep=1pt]
    \item Yes.
    \item No.
  \end{enumerate}
  
  \item Have you ever tried to invoke a skill during the interaction with another skill? (Except when you were listening to music)
  \begin{enumerate}[topsep=2pt,itemsep=1pt,parsep=1pt]
    \item Yes.
    \item No.
  \end{enumerate}
  
  \item Have you ever tried to turn up or turn down volume while interacting with a skill? (Except when you were listening to music)
  \begin{enumerate}[topsep=2pt,itemsep=1pt,parsep=1pt]
    \item Yes.
    \item No.
  \end{enumerate}
  
  \item What are the most frequent ways you have used to quit a skill? Please choose all that apply.
  \begin{checkbox}
    \item Alexa, stop.
    \item Alexa, cancel.
    \item Alexa, shut up.
    \item Alexa, cancel.
    \item Alexa, never mind.
    \item Alexa, forget it.
    \item Alexa, exit.
    \item Other (please specify).
  \end{checkbox}
  
  \item Have you ever experienced saying quit words (like the ones in the previous question) to a skill that you intended to quit but did not actually quit it?
  \begin{enumerate}[topsep=2pt,itemsep=1pt,parsep=1pt]
    \item Yes.
    \item No.
  \end{enumerate}
  
  \item Which indicator did you use most often to know that a conversation with Alexa is ended?
  \begin{enumerate}[topsep=2pt,itemsep=1pt,parsep=1pt]
    \item Alexa says ``Goodbye'', ``Have a good day'' or something similar.
    \item Alexa does not talk anymore.
    \item The light on the device is off.
    \item Other (please specify).
  \end{enumerate}
\end{enumerate}

\section{Context switch Examples}
\label{appendix:context_switch}

Here, we show some interesting examples of context switches discovered by the detector (Section~\ref{sec:defense}) in real world conversations collected by skills we published (see Section~\ref{subsec:real_world_attacks}). The examples presented here are transcripts including user utterances and their prior skill responses.

\vspace{3pt}\noindent\textbf{Skill}: Hello, welcome to soothing sleep sounds. Which sleep sound would you like today?

\vspace{3pt}\noindent\textbf{User utterances for context switch:}
\begin{itemize}[topsep=1pt]
\itemsep0em
\item Switch off the TV.
\item What time?
\item What is the week's forecast?
\item Show me the news.
\end{itemize}
\rule{\columnwidth}{0.5pt}

\vspace{3pt}\noindent\textbf{Skill}: Sorry, I do not understand. Which sound do you want today?

\vspace{3pt}\noindent\textbf{User utterances for context switch:}
\begin{itemize}[topsep=1pt]
\itemsep0em
\item Turn off Bluetooth.
\item Goodbye, Alexa.
\item I meant walk back to the timer.
\item Amazon music.
\item What's the weather in Northridge?
\item What's in the news?
\item I'm home.
\end{itemize}
\rule{\columnwidth}{0.5pt}

\vspace{3pt}\noindent\textbf{Skill}: Hello, welcome to my sleep sounds. Which sleep sound would you like today?

\vspace{3pt}\noindent\textbf{User utterances for context switch:}
\begin{itemize}[topsep=1pt]
\itemsep0em
\item Tell me a quote.
\item What was the time?
\end{itemize}
\rule{\columnwidth}{0.5pt}

\vspace{3pt}\noindent\textbf{Skill}: Hello, welcome to incredible fast sleep. Which sleep sound would you like today?

\vspace{3pt}\noindent\textbf{User utterances for context switch:}
\begin{itemize}[topsep=1pt]
\itemsep0em
\item What's my flash briefing?
\end{itemize}

\ignore{
\section{Cost matrix}
\label{appendix:cost_matrix}

Table \ref{tab:clip_matrix} shows some items of the cost matrix. \todo{add some explanations of the matxi}.

\begin{table}
\scriptsize
\centering
\caption{A clip of the pronunciation operation cost matrix}
\begin{tabular}{ccccccc}
\toprule
&AA&AE&N&CH&SH&none\\
\toprule
AA&N/A&0.7301&0.9981&1.0000&0.9982&0.9737\\
AE&0.7301&N/A&1.0000&1.0000&1.0000&0.9963\\
N&0.9981&1.0000&N/A&1.0000&1.0000&0.9774\\
CH&1.0000&1.0000&1.0000&N/A&0.4670&0.9936\\
SH&0.9982&1.0000&1.0000&0.4670&N/A&0.9986\\
none&0.9737&0.9963&0.9774&0.9936&0.9986&N/A\\
\bottomrule
\end{tabular}
\label{tab:clip_matrix}
\vspace{-15pt}
\end{table}
}

\end{appendices}